%% file: main.tex
\documentclass[a4paper,12pt]{article}
\usepackage[left=2.2cm,right=2.2cm,top=2cm,bottom=2.5cm]{geometry}

\usepackage[pdftex]{graphicx}
\graphicspath{{plots/}}
\usepackage{subcaption}

\usepackage{lscape} 
\usepackage{array}

\usepackage{fix-cm}

\usepackage[makeroom]{cancel}
\usepackage[normalem]{ulem}
\usepackage{amsmath,amssymb}
\usepackage{slashed}
\usepackage{cite}
\usepackage{pdflscape}
\usepackage{multirow}
\usepackage[thinlines]{easytable}
\usepackage{url}
\usepackage[utf8]{inputenc}
\usepackage{longtable}
\usepackage{booktabs}

\usepackage{braket}
\usepackage{siunitx}
\usepackage{verbatim}
\usepackage[dvipsnames]{xcolor}
\usepackage{xspace}
\usepackage{hyperref}
\usepackage{caption}

\usepackage{tocloft}
\newcommand{\scetone}{{\rm SCET}_{\rm I}}
\newcommand{\scettwo}{{\rm SCET}_{\rm II}}

\allowdisplaybreaks 

\input{macros.tex}

\begin{document}

\begin{titlepage}
\thispagestyle{empty}

\begin{center}
{\LARGE \textbf{SCET sum rules for $\Lambda_b$ $\rightarrow$ $\Lambda \ell^+\ell^-$,\,$\Lambda \gamma$ decays}} \\[1.5cm]

{\textbf{Long-Shun Lu}$^{a,b}$\footnote{E-mail: lulongshun@ihep.ac.cn}, \textbf{Cai-Dian Lü}$^{a,b}$, \textbf{Yue-Long Shen}$^{c}$, \textbf{Yan-Bing Wei}$^{d}$} \\[1cm]

{\sl   ${}^a$ \, Institute of High Energy Physics, CAS, P.O. Box 918(4) Beijing 100049,  China \\
 ${}^b$ \,  School of Physics, University of Chinese Academy of Sciences, Beijing 100049,  China \\
 ${}^c$ \, College of Information Science and Engineering,
Ocean University of China, Songling Road 238, Qingdao, 266100 Shandong, P.R. China \\
${}^d$ \,Institute of Particle Physics and Key Laboratory of Quark and Lepton Physics (MOE),
Central China Normal University, Wuhan, Hubei 430079, China
}
\end{center}

\vspace{0.2cm}
\begin{abstract}

We construct light-cone sum rules for various types of effective form factors in the $\Lambda_b \to \Lambda \ell^+\ell^-$ and $\Lambda_b \to \Lambda \gamma$ decays by analyzing vacuum-to-$\Lambda_b$ (or $\gamma^\ast$-to-$\Lambda_b$) correlation functions with the light $\Lambda$-baryon interpolating current. These form factors, defined via hadronic matrix elements within soft-collinear effective theory (SCET), enter the next-to-leading-power QCD factorization formulas for large-recoil transitions. 
Implementing the perturbative matching from $\text{SCET}_\text{I}$ to heavy quark effective theory, we determine the hard-collinear functions at next-to-leading-order accuracy. Based on light-cone sum rule predictions for the $\Lambda_b \to \Lambda$ form factors, we compute the $q^2$-dependent differential branching fraction, forward-backward asymmetry and dilepton longitudinal polarization fraction for $\Lambda_b \to \Lambda \ell^+\ell^-$ decay, as well as the branching fraction for $\Lambda_b \to \Lambda \gamma$ decay.

\end{abstract}
\end{titlepage}
\tableofcontents

\section{Introduction}

It is widely recognized that decays of bottom hadrons induced by the flavor-changing neutral current (FCNC) transition $b \to s\ell^+\ell^-$, with $\ell = e, \mu, \tau$, serve as sensitive probes of physics beyond the Standard Model (SM)~\cite{LHCb:2012myk,Belle-II:2018jsg}. Thanks to the large datasets collected at the Large Hadron Collider (LHC) and the Super B Factory, the properties of many exclusive $b \to s \ell^+\ell^-$ decay channels have been measured with unprecedented precision. The accumulation of additional data is expected to further enhance the statistical power of these measurements in the near future.
The measured observables in $b \to s \ell^+\ell^-$ decay providing valuable inputs for testing theoretical predictions and refining our understanding of rare bottom-hadron decays. Notably, tensions still exist between SM predictions and experimental measurements of these observables, particularly in the differential branching fractions and angular observables in the large-recoil region. 
Transition form factors, such as those in $B \to K^\ast$ and $B_s \to \phi$ decays, are essential nonperturbative inputs in exclusive $b \to s\ell^+\ell^-$ processes. These form factors have been carefully calculated using lattice QCD in the low-recoil region~\cite{Horgan:2013hoa,Horgan:2015vla}. In the large-recoil region, their predictions rely on extrapolation models, which can be improved by combining direct calculations with light-cone sum rules (LCSR)~\cite{Colangelo:2000dp,Khodjamirian:2023wol}.
Non-local hadronic matrix elements can also significantly affect the observables, and are  more challenging on the theoretical side. Although considerable effort has been devoted to evaluating such matrix elements, they continue to be a major source of theoretical uncertainty.

The baryon decay modes $\Lambda_b \to \Lambda \ell^+\ell^-$ offer new insights into the existing tensions between SM predictions and experimental data~\cite{LHCb:2018jna}. These decays also provide a rich set of angular observables that can help disentangle the contributions from different operators in the weak effective Hamiltonian. The decay $\Lambda_b \to \Lambda(\to p\pi) \ell^+\ell^-$ involves three angular variables, even when both the $\Lambda_b$ baryon and the proton are unpolarized~\cite{Gutsche:2013pp,Boer:2014kda}.
Moreover, since the $\Lambda$ is a ground-state baryon, the $\Lambda_b \to \Lambda$ transition form factors can be predicted with relatively higher accuracy, compared to those for $B \to K^\ast$ transitions. Similarly, the radiative decay $\Lambda_b \to \Lambda \gamma$ serves as a sensitive probe of the $b \to s \gamma$ transition, where the SM predicts predominantly left-handed photon polarization~\cite{LHCb:2019wwi}. The helicity of the final-state $\Lambda$ baryon is experimentally accessible, providing direct information about the photon polarization structure and enabling complementary constraints on possible right-handed currents beyond those observed in $B$-meson decays.
Taken together, these $\Lambda_b$ decay channels serve as golden modes for precision tests of the SM and for probing potential new physics scenarios.

The decay amplitudes of the $\Lambda_b \to \Lambda \ell^+\ell^-, \Lambda \gamma$ processes are determined by the matrix elements $\langle \ell^+\ell^-(\gamma)\Lambda |\mathcal{O}_i|\Lambda_b\rangle$, where $\mathcal{O}_i$ are the effective operators in the weak Hamiltonian. The most important hadronic inputs are the $\Lambda_b \to \Lambda$ transition form factors, which arise from the matrix elements of the semileptonic or electromagnetic operators. The $\Lambda_b \to \Lambda$ form factors in the small recoil region can be obtained from lattice-QCD studies~\cite{Detmold:2016pkz,Meinel:2023wyg}, and in the large recoil region one could employ LCSR to obtain relatively reliable predictions~\cite{Wang:2009hra,Feldmann:2011xf,Wang:2015ndk}.
In the heavy $b$-quark limit, the number of independent form factors is reduced from ten to two at small recoil, and large recoil symmetry further reduces this number to one~\cite{Feldmann:2011xf,Mannel:2011xg}. In the context of soft-collinear effective theory (SCET)~\cite{Bauer:2000yr,Bauer:2001yt,Beneke:2002ph}, this large recoil symmetry can be described in a more transparent and systematic way, allowing for an expansion of the form factors in powers of $\Lambda_{\rm QCD}/m_b$~\cite{Beneke:2003pa}. 
It has been established that the leading-power contribution originates from hard-collinear gluon exchange diagrams, which however is numerically subdominant compared to the soft contribution~\cite{Feldmann:2011xf}.
Beyond the local form factor contributions, the matrix elements of four-quark operators and the chromo-magnetic operator give rise to so-called “nonlocal form factors.” These include effects such as long-distance quark-loop contributions and the “annihilation topology contribution,” among others. In~\cite{Feldmann:2023plv}, the leading annihilation topology contribution was studied, revealing that it induces corrections of approximately 1\% relative to the local form factor contribution.

In this article, we perform a systematic analysis of the $\Lambda_b \to \Lambda \ell^+ \ell^-$ and $\Lambda_b \to \Lambda \gamma$ decays within the framework of SCET. Starting from the effective Hamiltonian for weak interactions in heavy quark decays, we match the effective operators onto the SCET$_{\rm I}$ operators. Since this work considers amplitudes only up to $\mathcal{O}(\alpha_s)$, we do not include SCET$_{\rm I}$ operators involving more than one gluon field.
The relevant SCET$_{\rm I}$ operators can be categorized into three types: A-type, B-type, and C-type. The decay amplitudes are then expressed as the product and convolution of effective hard functions and the matrix elements of SCET$_{\rm I}$ operators. We employ LCSR to evaluate the matrix elements of the A-type, B-type, and C-type operators, respectively. Among these, the A-type operators yield the dominant contributions.
Since the leading-power amplitude starts at $\mathcal{O}(\alpha_s^2)$, our analysis primarily focuses on the soft form factors that preserve large recoil symmetry. At the amplitude level, we find that the B-type operator contributions are strongly suppressed and can be neglected, while the non-factorizable C-type contributions are numerically insignificant. Consequently, the $\Lambda_b$ decay processes are dominated by soft contributions.

The remainder of this article is organized as follows. In the next section, we conduct a theoretical analysis of the $\Lambda_b \to \Lambda \ell^+ \ell^-$ and $\Lambda_b \to \Lambda \gamma$ decay amplitudes within the framework of SCET. In Section~\ref{section: SCET sum rules at leading twist}, we construct light-cone sum rules for the SCET $\Lambda_b \to \Lambda$ form factors, utilizing the distribution amplitude of the $\Lambda_b$ baryon. Numerical results are presented in Section~\ref{section: numerical analysis}. Section~\ref{section: phenomenologies} discusses several phenomenological applications. The final section concludes the article with a summary.

\section{SCET analysis for \texorpdfstring{$\Lambda_b \to \Lambda \ell^+ \ell^-, \Lambda \gamma$}{Lambda\_b → Lambda l+ l-, Lambda gamma} decays}
\label{section: SCET-I factorization}

In this section, we perform an analysis of the $\Lambda_b \to \Lambda \ell^+ \ell^-$ and $\Lambda_b \to \Lambda \gamma$ decays within the framework of SCET.  
In the rest frame of the $\Lambda_b$ baryon, the final-state $\Lambda$ baryon moves at high velocity, and we assume that the momenta of the quarks inside the $\Lambda$ baryon are aligned along a light-cone direction $\bar{n}$.
For the $\Lambda_b \to \Lambda \gamma$ decay, the momentum of the photon is given by $q_\gamma = E_\gamma n$, with ${\bar n} \cdot n = 2$.
For the $\Lambda_b \to \Lambda \ell^+ \ell^-$ decays, we restrict our analysis to the kinematic region $1~\mbox{GeV}^2 < q^2 < 7~\mbox{GeV}^2$, where $q$ denotes the momentum of the lepton pair. In this region, resonance effects can be safely neglected. 

A given momentum $p$ can be decomposed as 
$p^\mu= n \cdot p \, {\bar n}^\mu/2+ {\bar n}\cdot p \, n^\mu/2+ p_\perp^\mu $.
The momentum of the $\Lambda$ baryon follows the power-counting rule
\begin{equation}
	p = (n \cdot p, \bar{n} \cdot p, p_\perp) \sim (1, \lambda^4, \lambda^2)\, m_b~,
\end{equation}
with $\lambda \sim \sqrt{\Lambda_{\rm QCD}/m_b}$. In addition to this collinear momentum mode, the power counting for the soft and hard-collinear modes is $(\lambda^2, \lambda^2, \lambda^2)\, m_b$ and $(1, \lambda^2, \lambda)\, m_b$, respectively. 
Moreover, the anti-collinear photon and the anti-hard-collinear lepton pair scale as $q_\gamma \sim (\lambda^4, 1, \lambda^2)\, m_b$ and $q \sim (\lambda^2, 1, \lambda)\, m_b$, respectively.
The fields in SCET, in terms of gauge invariant building blocks~\cite{Hill:2002vw}, also have definite power counting
\begin{eqnarray}
	&&\xi_q \sim \lambda~,\quad
 \chi_q \sim \lambda~,\quad A_{hc}^\mu \sim (1, \lambda^2, \lambda)~,\quad A_{\overline{hc} \perp}^{(\rm em)} \sim \lambda~,\quad q_{q} \sim \lambda^{3}~,
\quad h \sim \lambda^3. 
\end{eqnarray}
Above, the symbols $\xi_q$, $\chi_q$, $q_q$ and $h$ stand for the  hard-collinear, anti-hard-collinear, soft quark and effective heavy quark fields, respectively, where the subscript $q = u, d, s$ denotes the flavor of the light quarks. The symbols $A_{hc}^\mu$ and $A_{\overline{hc} \perp}^{(\rm em)}$ stand for hard-collinear gluon field and  anti-hard-collinear photon field, separately.

Since there exist both hard-collinear and collinear fields, two effective theories are introduced. The first is an intermediate effective theory, denoted as $\scetone$, which contains soft and hard-collinear degrees of freedom. The second is the final infrared theory, referred to as $\scettwo$, which contains soft and collinear fields. 
For a factorizable process, long-distance nonperturbative quantities, such as the light-cone distribution amplitudes (LCDAs) of the initial- and final-state hadrons, are described by the matrix elements of $\scettwo$ operators. The perturbative functions can be computed through a two-step matching procedure: QCD $\to$ $\scetone$ $\to$ $\scettwo$.

\subsection{The operators in SCET}
The effective weak Hamiltonian for the semileptonic $b \to s \ell^{+} \ell^{-}, s \gamma$ transitions in the SM can be written as \cite{Beneke:2004dp}
\begin{equation}
	{\cal H}_{\rm eff} =  - \frac{G_F}{\sqrt{2}}\, \left[ V_{tb} V_{ts}^{\ast} \,  {\cal H}_{\rm eff}^{(t)}
	+ V_{ub} V_{us}^{\ast} \,  {\cal H}_{\rm eff}^{(u)} \right] + {\rm h.c.}   \,,
	\label{effective weak Hamiltonian of b to s}
\end{equation}
where we have employed the unitarity relations of the Cabibbo-Kobayashi-Maskawa (CKM) matrix elements, and
\begin{align}
	{\cal H}_{\rm eff}^{(t)} &= C_1 \, {\cal Q}_1^{c} + C_2 \, {\cal Q}_2^{c} + \sum_{i=3,\ldots,10} C_i \, {\cal Q}_i \,, \nonumber \\
	{\cal H}_{\rm eff}^{(u)} &= C_1 \, ({\cal Q}_1^{c} - {\cal Q}_1^{u})  + C_2 \, ({\cal Q}_2^{c} - {\cal Q}_2^{u})  \,.
\end{align}
Since ${\cal H}_{\rm eff}^{(u)}$ is doubly Cabibbo-suppressed, it can be safely neglected. In full QCD, the hadronic matrix elements of the semileptonic operators ${\cal Q}_{7, 9, 10}$ can be readily expressed in terms of the heavy-to-light $\Lambda_b \to \Lambda$ decay form factors \cite{Feldmann:2011xf,Wang:2015ndk}. The remaining contributions to the electroweak penguin decay amplitude can be determined by contracting the exclusive $\Lambda_b \to \Lambda \, \gamma^{\ast}$ matrix element with the vector lepton current $\gamma^{\ast} \to \ell^{+} \ell^{-}$ in the $\Lambda_b \to \Lambda \ell^+ \ell^-$ decay.

The matching from QCD to $\scetone$ can be expressed as
\begin{align}
{\cal H}_{b \to s \ell^+ \ell^-} \, &= \,
\sum_{i=1}^4 \int \! ds~\widetilde{C}_i^A (s) Q_i^A (s) 
+ \sum_{j=1}^4 \int \! ds \int \! dr~\widetilde{C}_j^B (s,r) Q_j^B (s,r)  \nonumber \\
&\quad + \sum_{k=1}^4 \int \! ds \int \! dr \int \! dt~\widetilde{C}_k^C(s,r,t) Q_k^C(s,r,t) + \dots~, \\
{\cal H}_{b \to s \gamma} \, &= \,
\int \! ds \int \! dt~\widetilde{C}_\gamma^A (s,t) Q_\gamma^A (s,t) 
+ \sum_{j=1}^2 \int \! ds \int \! dr \int \! dt~\widetilde{C}_{\gamma,j}^B (s,r,t) Q_{\gamma,j}^B (s,r,t)  \nonumber \\
&\quad + \sum_{k=1}^4 \int \! ds \int \! dr \int \! dt~\widetilde{C}_k^C(s,r,t) Q_k^C(s,r,t) + \dots~,
\label{eq:factorization.operator}
\end{align}
where $\widetilde{C}_i^{(A,B,C)}$, $\widetilde{C}_{\gamma}^A$, and $\widetilde{C}_{\gamma,i}^B$ are hard functions in position space. These are related to the momentum-space hard functions via the Fourier transformations:
\begin{align}
	C_{i}^A(n\cdot p) &= \int \! ds~e^{isn\cdot p} \, \widetilde{C}_{i}^A(s)~, \nonumber \\
	C_{i}^B(n\cdot p,u) &= \int \! ds \int \! dr~e^{i(us+{\bar u}r)n\cdot p} \, \widetilde{C}_{i}^B(s,r)~,\nonumber \\
	C_{\gamma}^A(n\cdot p) &= \int \! ds \int \! dt~e^{isn\cdot p} e^{it \bar{n}\cdot q} \, \widetilde{C}_{\gamma}^A(s,t)~, \nonumber \\
	C_{\gamma,i}^B(n\cdot p,u) &= \int \! ds \int \! dr \int \! dt~e^{i(us+{\bar u}r)n\cdot p} e^{it \bar{n}\cdot q} \, \widetilde{C}_{\gamma,i}^B(s,r,t)~, \nonumber \\
	C_{i}^C(n\cdot p,u) &= \int \! ds \int \! dr \int \! dt~e^{i(us+{\bar u}r)n\cdot p} e^{it \bar{n}\cdot q} \, \widetilde{C}_i^C(s,r,t)~,
\end{align}
with ${\bar u} = 1 - u$. 

The A-type and B-type $\scetone$ operators for $b \to s \ell^+ \ell^-$ decay are given by \cite{Ali:2006ew}:
\begin{equation}
\begin{aligned}
Q_{1(3)}^A &= \bar{\xi}_{s}(sn)(1+\gamma_5)\gamma_\perp^\mu h(0)~ \bar{\ell} \gamma_\mu(\gamma_\mu\gamma_5) \ell~, \\
Q_{2(4)}^A &= \bar{\xi}_{s}(sn)(1+\gamma_5) \frac{\bar{n}^\mu}{\bar{n}\cdot v} h(0)~ \bar{\ell} \gamma_\mu(\gamma_\mu\gamma_5) \ell~, \\
Q_{1(3)}^B &= \bar{\xi}_{s}(sn)(1+\gamma_5)\gamma_\perp^\mu \slashed{A}_{hc\perp}(rn) h(0)~ \bar{\ell} \gamma_\mu(\gamma_\mu\gamma_5) \ell~, \\
Q_{2(4)}^B &= \bar{\xi}_{s}(sn)(1+\gamma_5)\slashed{A}_{hc\perp}(rn) \frac{\bar{n}^\mu}{\bar{n}\cdot v} h(0)~ \bar{\ell} \gamma_\mu(\gamma_\mu\gamma_5) \ell~, \\
\end{aligned}
\end{equation}
and the operators for $b \to s \gamma$ decay are \cite{Becher:2005fg}:
\begin{equation}
\begin{aligned}
Q_\gamma^A &= \bar{\xi}_{s}(sn)(1+\gamma_5)\slashed{A}^{\rm (em)}_{\overline{hc} \perp}(t \bar n) h(0)~, \\
Q_{\gamma,1}^B &= \bar{\xi}_{s}(sn)(1+\gamma_5)\slashed{A}^{\rm (em)}_{\overline{hc} \perp}(t \bar n) \slashed{A}_{hc\perp}(rn) h(0)~, \\
Q_{\gamma,2}^B &= \bar{\xi}_{s}(sn)(1+\gamma_5)\slashed{A}_{hc\perp}(rn) \slashed{A}^{\rm (em)}_{\overline{hc} \perp}(t \bar n) h(0)~.
\end{aligned}
\end{equation}
The A-type and B-type operators describe the emission of the (virtual) photon from the $b \to s$ transition currents, while the C-type operators $Q_i^C$ correspond to diagrams in which the (virtual) photon is emitted from the spectator quark in the $\Lambda_b$ baryon. These represent the non-factorizable contributions of the strong-penguin operators \cite{Feldmann:2023plv}. For both $\Lambda_b \to \Lambda \ell^+ \ell^-$ and $\Lambda_b \to \Lambda \gamma$ decays, the relevant C-type $\scetone$ operators are:
\begin{equation}
\begin{aligned}
Q_{1,3}^C &= \bar{\xi}_{s}(sn)(1+\gamma_5)\gamma_\perp^\mu h(0)~ \bar{\xi}_{q}(rn)\gamma^\perp_\mu(1\mp\gamma_5) \chi_q(t \bar n)~, \\
Q_{2,4}^C &= \bar{\xi}_{s}^i(sn)(1+\gamma_5)\gamma_\perp^\mu h^j(0)~ \bar{\xi}_{q}^j(rn)\gamma^\perp_\mu(1\mp\gamma_5) \chi_q^i(t \bar n)~, \\
\end{aligned}
\end{equation}
where the subscript $q = u, d$ denotes the spectator quark that emits the (virtual) photon. 

\begin{figure}[t]
\centering
\includegraphics[width=.9 \columnwidth]{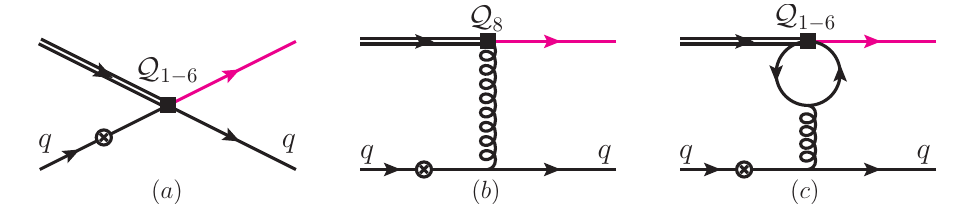}
\caption{Feynman diagrams where the (virtual) photon, as denoted by the crossed circle, is emitted from
the spectator light quark $q$. The $b$-quark is indicated by a double line; the strange quark by a magenta line.}
\label{fig: Ctypematching}
\end{figure}

The hard functions of A-type operators have been calculated up to the two-loop level, except for the contributions from quark-loop diagrams with insertions of penguin operators. The coefficients of B-type operators are known at the one-loop level. In this work, we aim at a one-loop calculation of the decay amplitude. Therefore, we utilize the existing one-loop results for  ${C}_i^{(A,B)}$ from \cite{Ali:2006ew}, as well as ${C}^{A}_{\gamma}$ and ${C}^{B}_{\gamma,i}$ from \cite{Becher:2005fg}. 
To obtain the hard functions of C-type operators, we consider the diagrams shown in Fig.~\ref{fig: Ctypematching}, where the photon is emitted from the spectator quark. The calculation of the hard function up to $\alpha_s$-level is analogous to the annihilation contribution to the $B\to \{\pi, K\}\ell^+\ell^-$ decays \cite{Huang:2024xii}.  We use the notation $\Delta_{16}^{(0)}C_i^C$, $\Delta_{12}^{(1)}C_i^C$ and $\Delta_{8}^{(1)}C_i^C$ stand
for the matching of ${\cal Q}_{1-6}$ and ${\cal Q}_8$ onto the C-type $\scetone$ operators $C_i^C$ respectively. The annihilation-type leading topology is illustrated in diagram (a) of Fig.~\ref{fig: Ctypematching}, contributes to the matching coefficient $C_i^C$ at order $\alpha_s^0$, for which the
calculation is trivial:
\begin{equation}
\Delta_{16}^{(0)}C^C_{1,2,3,4} = -\frac{G_F}{\sqrt{2}}\, V_{t b} V_{t s}^{\ast} \, \bar{C}_{3,4,5,6} \,,
\end{equation}
where the "barred" coefficients $\bar{C}_{1-6}$ are
related to the Wilson coefficients in the standard BBL basis~\cite{Buchalla:1995vs} and the superscript (0) denotes the matching at order $\alpha_s^0$. The chromomagnetic contribution is originate from chromomagnetic operator ${\cal Q}_{8}$ at order $\alpha_s$, as shown in diagram (b) of Fig.~\ref{fig: Ctypematching}, matching onto the C-type $\scetone$ operators with the coefficients
\begin{align}
\Delta_{8}^{(1)}C^C_{1} &= \Delta_{8}^{(1)}C^C_{3} = -\frac{G_F}{\sqrt{2}} \, V_{t b} V_{t s}^{\ast} \, \frac{1}{N_c} \, \frac{\alpha_s}{4 \pi} \, \frac{C_8^{\rm eff}}{1-u+u\hat{s}} \,, \nonumber \\
\Delta_{8}^{(1)}C^C_{2} &= \Delta_{8}^{(1)}C^C_{4} = \frac{G_F}{\sqrt{2}} \, V_{t b} V_{t s}^{\ast} \, \frac{\alpha_s}{4 \pi} \, \frac{C_8^{\rm eff}}{1-u+u\hat{s}} \,,
\end{align}
where $\hat{s} \equiv q^2/m_b^2$, $C_8^{\rm eff}=C_8+(4\bar C_3-\bar C_5)/3$~\cite{Beneke:2001at}. Diagram (c) in Fig.~\ref{fig: Ctypematching} corresponds to the quark loops with spectator scattering (QLSS), which arise from the operators ${\cal Q}_{1-6}$ at $\mathcal{O}(\alpha_s)$. The dominant contribution comes from the charm quark loop induced by ${\cal Q}_2$ due to its large Wilson coefficient. Since the penguin operators yield negligible effects, we omit their contributions. Thus, this diagram yields the following matching coefficients:
\begin{align}
\Delta_{12}^{(1)}C^C_{1} = \Delta_{12}^{(1)}C^C_{3} = & \,\frac{G_F}{\sqrt{2}} \, V_{t b} V_{t s}^{\ast} \, \frac{1}{2 N_c} \, \frac{\alpha_s}{4 \pi} \, \bar C_2 \, G\left(u,\hat{s},m_c^2/m_b^2\right)\,, \nonumber \\
\Delta_{12}^{(1)}C^C_{2} = \Delta_{12}^{(1)}C^C_{4} = & -N_c \, \Delta_{12}^{(1)}C^C_{1} = -N_c \, \Delta_{12}^{(1)}C^C_{3}\,,
\end{align}
where the function $G\left(u,\hat{s},\lambda \right)$ is defined as
\begin{align}
    G\left(u,\hat{s},\lambda \right) = \frac{2}{3} + \frac{2}{3} \ln{\frac{m_b^2}{\mu^2}} + 4\int_0^1 dx \, x(1-x)\ln{\left[\lambda - x(1-x)(1-u+u\hat{s})\right]}\,.
\end{align}

These subleading topologies from diagrams (b) and (c) in Fig.~\ref{fig: Ctypematching} are indeed numerically suppressed.  Meanwhile, it is important to note that the contribution from diagram (c) of Fig.~\ref{fig: Ctypematching} is essential for eliminating the infrared divergence originating from the vertex correction diagrams in diagram (a) of Fig.~\ref{fig: Ctypematching}. Finally, up to order $\alpha_s$, we have:
\begin{equation}
C_i^C = \Delta_{16}^{(0)}C^C_i + \Delta_{8}^{(1)}C^C_i + \Delta_{12}^{(1)}C^C_i\,.
\end{equation}
It is worth noting that we neglect the ``vertex corrections" to the weak annihilation graph, and the matrix elements $\langle {Q}_i^C \rangle$ are computed only at tree level, given the the small Wilson coefficients of QCD penguin operators and the strong coupling constant. Consequently, the matching remains incomplete at order $\alpha_s$.

\subsection{The decay amplitudes in SCET}

At the amplitude level, we consider standard relativistic normalization for hadronic states and have
\begin{align}
&\, \langle \Lambda(p,s') \, \ell^-(p_1,s_1) \, \ell^+(p_2,s_2)| -i\,{\cal H}_{b \to s \ell^+ \ell^-} | \Lambda_b(v,s) \rangle 
\nonumber \\
 =& \, (2\pi)^{4} \delta^{4}(M_{\Lambda_b}v-p-p_1-p_2) \, i\mathcal{M}(s,s',s_1,s_2)\,, 
\end{align}
\begin{align}
\langle \Lambda(p,s') \, \gamma(q,\lambda)| -i\, {\cal H}_{b \to s \gamma}\, | \Lambda_b(v,s) \rangle = \, (2\pi)^{4} \delta^{4}(M_{\Lambda_b}v-p-q) \, i\mathcal{M}(s,s',\lambda)\,,
\end{align}
where $\mathcal{M}(s,s',s_1,s_2)$ and $\mathcal{M}(s,s',\lambda)$ are the helicity amplitudes for the decays $\Lambda_b \to \Lambda \ell^+ \ell^-$ and $\Lambda_b \to \Lambda  \gamma$, respectively. To obtain the factorized helicity amplitudes from \refeq{factorization.operator}, we must appropriately define form factors for the matrix elements of various $\scetone$ operators. 

As noted in the introduction, within the heavy quark limit and large recoil limit, there exists a unique form factor  $\xi_\Lambda(n\cdot p)$, which   correspond to the matrix elements of A-type $\scetone$ operators\cite{Feldmann:2011xf,Mannel:2011xg}
\begin{equation}
\langle \Lambda(p,s')| \bar{\xi}_{s}(0) \, \Gamma_i^A \, h(0) | \Lambda_b(v,s) \rangle 
 = \, \xi_\Lambda(n \cdot p) \, \bar u_\Lambda(p,s') \, \Gamma_i^A \, u_{\Lambda_b}(v,s)\,.
  \end{equation}
$\xi_\Lambda(n\cdot p)$ is assumed
to be dominated by the soft gluon exchanges. 

 The matrix elements of B-type $\scetone$ operators correspond to hard-scattering corrections from hard-collinear gluon exchange~\cite{Feldmann:2011xf}, similar to the case of the $B \to P$ transition~\cite{Beneke:2000wa,DeFazio:2007hw,DeFazio:2005dx}. There is only one independent non-local form factor for B-type matrix elements, which we define as
\begin{align}
& \, n \cdot p \int \! \frac{da}{2\pi}~e^{-iu\, n \cdot p\, a}\, \langle \Lambda(p,s')| \, \bar{\xi}_{s}(0)\, \Gamma_i^B \, {A}_{hc\perp}^{\nu}(an) \, h(0) \, | \Lambda_b(v,s) \rangle \nonumber \\
= & \, M_{\Lambda_b} \, \Delta\xi_\Lambda^B(u,n \cdot p) \, \bar u_\Lambda(p,s') \, \gamma_{\perp}^{\nu} \, \Gamma_i^B \, u_{\Lambda_b}(v,s) \,.
\end{align}

 The C-type contribution is referred to as a ``non-factorisable contribution''~\cite{Feldmann:2023plv,Beneke:2001at,Wang:2015ndk,Wang:2014jya}. For the decay $\Lambda_b \to \Lambda \ell^+ \ell^-$, the lepton and hadronic parts can be factorized at the amplitude level:
\begin{align}
    & \, n \cdot p \int \! \frac{da}{2\pi}~e^{-iu\, n \cdot p\, a}\, \langle \Lambda(p,s') \, \ell^-(p_1,s_1) \, \ell^+(p_2,s_2)|\,{Q}_{i}^C(0,a,0) \,| \Lambda_b(v,s) \rangle \nonumber \\
    = \, & g_{\rm em}\, \frac{1}{q^2}\, \sum_\lambda \bar{\epsilon}^\mu(q,\lambda)\,  \bar u(p_1,s_1) \, \gamma_\mu \, v(p_2,s_2) \nonumber \\
    & \qquad \qquad \times \, n \cdot p \int \! \frac{da}{2\pi}~e^{-iu\, n \cdot p\, a}\, \langle \Lambda(p,s')\, \gamma^\ast(q,\lambda)|\,{Q}_{i}^C(0,a,0) \,| \Lambda_b(v,s) \rangle \,,
\end{align}
where $\bar{\epsilon}^\mu(q,\lambda)$ denotes the polarization vector of the virtual photon. The hadronic matrix elements of ${Q}_i^C$ are parametrized in terms of form factors as follows:
\begin{align}
    & \, n \cdot p \int \! \frac{da}{2\pi}~e^{-iu\, n \cdot p\, a}\, \langle \Lambda(p,s')\, \gamma^\ast(q,\lambda)|\,{Q}_{i}^C(0,a,0) \,| \Lambda_b(v,s) \rangle \nonumber \\
    =\, & g_{\rm em} \, M_{\Lambda_b}^2 \, \bar{\epsilon}_\nu^\ast(q,\lambda)\, \Delta_i\xi_\Lambda^C(u,n \cdot p) \, \bar u_\Lambda(p,s') \, (1+\gamma_5)\, \gamma_{\perp}^{\nu} \, u_{\Lambda_b}(v,s) \,.
\end{align}
In the decay $\Lambda_b \to \Lambda \gamma$ we have
\begin{align}
    & \, n \cdot p \int \! \frac{da}{2\pi}~e^{-iu\, n \cdot p\, a}\, \langle \Lambda(p,s')\, \gamma(q,\lambda)|\,{Q}_{i}^C(0,a,0) \,| \Lambda_b(v,s) \rangle \nonumber \\
    =\, & g_{\rm em} \, M_{\Lambda_b}^2 \, {\epsilon}_\nu^\ast(q,\lambda)\, \Delta_i\xi_\Lambda^C(u,n \cdot p) \, \bar u_\Lambda(p,s') \, (1+\gamma_5)\, \gamma_{\perp}^{\nu} \, u_{\Lambda_b}(v,s) \,,
\end{align}
where ${\epsilon}_\nu^\ast(q,\lambda)$ denotes the polarization vector of the real photon.

Using the definitions of the $\scetone$ form factors, we can derive the factorized form of the helicity amplitudes. For the $\Lambda_b \to \Lambda \ell^+ \ell^-$ decay, we have
\begin{align}
\mathcal{M}&(s,s',s_1,s_2) = - \sum_{i=1}^4 \Bigg\{
C_i^A(n \cdot p) \, \xi_\Lambda(n \cdot p) \, \bar u_\Lambda(p,s') \, \Gamma_i^A \, u_{\Lambda_b}(v,s) \,
\bar u(p_1,s_1) \, \Gamma_i^{\ell \ell} \, v(p_2,s_2)
\nonumber \\
&\quad + M_{\Lambda_b} \int \! du \, C_i^B(n \cdot p,\bar{u}) \, \Delta\xi_\Lambda^B(u,n \cdot p) \, \bar u_\Lambda(p,s') \, \gamma_{\perp}^{\nu} \, \Gamma_i^B \, u_{\Lambda_b}(v,s) \,
\bar u(p_1,s_1) \, \Gamma_i^{\ell \ell} \, v(p_2,s_2)
\nonumber \\
&\quad - 16 \pi^2 \, \frac{\alpha_{\rm e}}{4\pi} \, \frac{1}{q^2} \, M_{\Lambda_b}^2 \int \! du \, C_i^C(n \cdot p,\bar{u}) \, \Delta_i\xi_\Lambda^C(u,n \cdot p) \,
\bar u_\Lambda(p,s') \, (1+\gamma_5) \gamma_{\perp}^{\mu} \, u_{\Lambda_b}(v,s)
\nonumber \\
&\qquad \times \bar u(p_1,s_1) \, \gamma_\mu \, v(p_2,s_2) \Bigg\} \,.
\label{eq: SCET-I factorization formulae}
\end{align}
For the $\Lambda_b \to \Lambda \gamma$ decay, we have
\begin{align}
\mathcal{M}&(s,s',\lambda) = -\, \epsilon_\mu^*(q,\lambda) \Bigg\{
C^A_{\gamma}(n \cdot p)\, \xi_{\Lambda}(q^2=0)\, \bar u_\Lambda(p,s') \, (1+\gamma_5)\, \gamma_{\perp}^{\mu} \, u_{\Lambda_b}(v,s)
\nonumber \\
&\quad + M_{\Lambda_b} \sum_{j=1}^2 \int \! du \, C_{\gamma,j}^B(n \cdot p,\bar{u}) \, \Delta\xi_\Lambda^B(u,q^2=0) \,
\bar u_\Lambda(p,s') \, \gamma_{\perp}^{\nu} \, \Gamma_{\gamma,j}^B \, u_{\Lambda_b}(v,s)
\nonumber \\
&\quad + g_{\rm em} \, M_{\Lambda_b}^2 \sum_{k=1}^4 \int \! du \, C_k^C(n \cdot p,\bar{u}) \, \Delta_k\xi_\Lambda^C(u,q^2=0) \,
\bar u_\Lambda(p,s') \, (1+\gamma_5) \gamma_{\perp}^{\mu} \, u_{\Lambda_b}(v,s) \Bigg\} \,.
\label{eq:gamma factorization formulae}
\end{align}
The Dirac structures are defined as
\begin{equation}
\begin{aligned}
\Gamma_i^A &=
\begin{cases}
(1+\gamma_5)\gamma_\perp^\mu & i=1,3 \\
(1+\gamma_5)\, \dfrac{\bar{n}^\mu}{\bar{n} \cdot v} & i=2,4
\end{cases} \,, \qquad
\Gamma_i^{\ell \ell} =
\begin{cases}
\gamma_\mu & i=1,2 \\
\gamma_\mu \gamma_5 & i=3,4
\end{cases} \,, \\
\Gamma_i^B &=
\begin{cases}
(1+\gamma_5)\gamma_\perp^\mu \gamma_{\nu} & i=1,3 \\
(1+\gamma_5)\, \dfrac{\bar{n}^\mu}{\bar{n} \cdot v} \gamma_{\nu} & i=2,4
\end{cases} \,, \qquad
\Gamma_{\gamma,j}^B =
\begin{cases}
(1+\gamma_5)\gamma_\perp^\mu \gamma_{\nu} & j=1 \\
(1+\gamma_5)\gamma_{\nu} \gamma_\perp^\mu & j=2
\end{cases} \,.
\end{aligned}
\end{equation}

As shown above, due to the identical spinor structures, the non-factorisable contributions to the $\Lambda_b \to \Lambda \ell^+ \ell^-$ and $\Lambda_b \to \Lambda \gamma$ decay amplitudes can be absorbed into the definitions of $C_1^A$ and $C_\gamma^A$:
\begin{equation}
    C_1^A \to C_1^A + \Delta C_1^A(q^2) \,, \qquad
    C_\gamma^A \to C_\gamma^A + \Delta C_\gamma^A(q^2=0) \,,
\end{equation}
where
\begin{align}
    &\Delta C_1^A(q^2) = - 16 \pi^2 \, \frac{\alpha_{\rm e}}{4\pi} \, \frac{1}{q^2} \, M_{\Lambda_b}^2 \sum_{i=1}^4 \int \! du \, C_i^C(n \cdot p,\bar{u}) \, \frac{\Delta_i\xi_\Lambda^C(u,n \cdot p)}{\xi_\Lambda(n \cdot p)} \,, \nonumber \\
    &\Delta C_\gamma^A(q^2=0) = g_{\rm em} \, M_{\Lambda_b}^2 \sum_{i=1}^4 \int \! du \, C_i^C(n \cdot p,\bar{u}) \, \frac{\Delta_i\xi_\Lambda^C(u,q^2=0)}{\xi_\Lambda(q^2=0)} \,.
    \label{eq:deltaCA}
\end{align}

\section{LCSR for the SCET form factors}
\label{section: SCET sum rules at leading twist}

In this section, we construct the light-cone sum rules for the effective form factors
$\xi_\Lambda(n \cdot p)$, $\Delta\xi_\Lambda^B(u,n \cdot p)$, and $\Delta_i\xi_\Lambda^C(u,n \cdot p)$ appearing in the factorization formulae \refeq{ SCET-I factorization formulae} and \refeq{gamma factorization formulae}. We compute $\xi_\Lambda(n \cdot p)$ and $\Delta\xi_\Lambda^B(u,n \cdot p)$ at one-loop accuracy, while $\Delta_i\xi_\Lambda^C(u,n \cdot p)$ is evaluated at tree level.
To this end, we first derive the soft-collinear factorization theorems for the vacuum-to-$\Lambda_b$-baryon correlation functions, using an interpolating current for the collinear $\Lambda$-baryon in combination with A-type or B-type $\scetone$ operators. The resummation of parametrically large logarithms $\ln (m_b / \Lambda_{\rm QCD})$ is performed at leading logarithmic (LL) accuracy using the renormalization group (RG) equation formalism.
Subsequently, we extend this framework to derive the soft-collinear factorization theorems for the $\gamma^\ast$-to-$\Lambda_b$-baryon correlation functions. In this case, the C-type $\scetone$ operators, combined with the collinear $\Lambda$-baryon interpolating current, yield the form factors $\Delta_i\xi_\Lambda^C(u,n \cdot p)$.

We begin with a correlation function, where the $\Lambda$ baryon in the final state is replaced by an interpolating current that carries the same quantum numbers. We choose
\begin{align} 
J_{\Lambda}(x) = \epsilon_{ijk} \left[ u^{\rm T, i}(x) \, C \, \gamma_5 \, \slashed{n} \, d^j(x) \right] s^k(x) \,,
\end{align}
which is normalized via the matrix element
\begin{align}
\langle 0 | J_{\Lambda}(0) | \Lambda(p, s') \rangle = f_{\Lambda}(\mu) \,
(n \cdot p) \, u_\Lambda(p,s') \,.
\end{align}
To match the above current onto SCET, the light-quark fields are decomposed into soft and hard-collinear components. The SCET representation of the $\Lambda$ interpolating current can then be obtained following the prescriptions given in \cite{Beneke:2002ph}:
\begin{eqnarray}
J_{\Lambda} = J_{\Lambda}^{(0)} + J_{\Lambda}^{(2)} + J_{\Lambda}^{(4)} + \dots \,,
\end{eqnarray}
where the explicit expressions for the effective currents (noting that there is no soft $s$-quark field) are given by
\begin{align}
J_{\Lambda}^{(0)} =&\; \epsilon_{ijk} \left[ \xi_u^{\rm T, i} \, C \, \gamma_5 \, \slashed{n} \, \xi_d^j \right] \xi_s^k \,, \nonumber \\
J_{\Lambda}^{(2)} =&\; \epsilon_{ijk} \left[ q_u^{\rm T, i} \, C \, \gamma_5 \, \slashed{n} \, \xi_d^j \right] \xi_s^k
- \epsilon_{ijk} \left[ q_d^{\rm T, i} \, C \, \gamma_5 \, \slashed{n} \, \xi_u^j \right] \xi_s^k \,, \nonumber \\
J_{\Lambda}^{(4)} =&\; \epsilon_{ijk} \left[ q_u^{\rm T, i} \, C \, \gamma_5 \, \slashed{n} \, q_d^j \right] \xi_s^k \,.
\end{align}

We perform the calculations in the light-cone gauge for both hard-collinear and soft gluon fields. The hard-collinear gluon propagator is given by
\begin{align}
\langle 0| T \left[ A_{hc, \mu}^a(x) \, A_{hc, \nu}^b(y) \right] | 0 \rangle 
&= \int \frac{d^D l}{(2 \pi)^D} \, \frac{-i \, \delta^{ab} }{l^2 + i\epsilon} \, e^{-i l \cdot (x - y)} 
\left[ g_{\mu \nu} - \frac{1}{n \cdot l} \, (l_{\mu} \, n_{\nu} + n_{\mu} \, l_{\nu}) \right] \,.
\end{align}
The multipole-expanded SCET Lagrangian up to ${\cal O}(\lambda^2)$ accuracy \cite{Beneke:2002ni} has been derived using the position-space formalism \cite{Beneke:2002ph}.
To ensure that the correlation function admits a light-cone operator-product expansion (OPE), we assign $p_\mu$ as a hard-collinear momentum with
\begin{equation}
n \cdot p \simeq \frac{M_{\Lambda_b}^2 - q^2}{M_{\Lambda_b}} \simeq 2E_\Lambda \,, \quad
| \bar n \cdot p | \sim \mathcal{O}(\Lambda_{\rm QCD}) \,, \qquad p^2 < 0 \,.
\end{equation}

\subsection{LCSR for A-type soft form factor}

We now define the correlation function between the A-type weak decay current and
the interpolating current $J_\Lambda$, 
\begin{align}
 \Pi^{\rm A}(\bar{n} \cdot p) & \equiv i \int d^4x \, e^{ip \cdot x} \langle 0| T \left[ J_\Lambda(x), O_i^A \right] |\Lambda_b(v,s)\rangle \,,
\end{align}
where $O_i^A = \bar{\xi}_{s}(0) \, \Gamma_i^A \, h(0)$. On the hadronic side, this correlation function is evaluated by inserting a complete set of intermediate states with the same quantum numbers as the $\Lambda$ baryon:
\begin{align}
 \Pi^A \big|_{\rm res.} & = \sum_{s'} 
   \frac{ 
\langle 0| J_\Lambda | \Lambda(p,s')\rangle
\langle \Lambda(p,s')| O_i^A | \Lambda_b(v, s) \rangle }{m_\Lambda^2 - p^2} + \cdots
\cr 
 &= \frac{(n \cdot p) \, f_\Lambda (\nu) \, \xi_\Lambda(n \cdot p)}{m_\Lambda^2 - (n \cdot p)(\bar{n} \cdot p)}
 \, \sum_{s'} u_\Lambda(p,s') \,
 \bar u_\Lambda(p,s') \, \Gamma_i^A \, u_{\Lambda_b}(v,s) + \cdots
\cr 
&= \frac{(n \cdot p) \, f_\Lambda (\nu) \, \xi_\Lambda(n \cdot p)}{m_\Lambda^2 / (n \cdot p) - (\bar{n} \cdot p)} \, \frac{\slashed {\bar n}}{2} \, \Gamma_i^A \, u_{\Lambda_b}(v,s) + \cdots \,,
\end{align}
where the ellipsis denotes contributions from excited and continuum states.

On the other hand, within our selected momentum region, where the light-cone operator-product expansion is applicable, the time-ordered product of the two currents can be computed using perturbation theory. In SCET terminology, by matching from $\scetone$ to HQET \cite{DeFazio:2005dx, DeFazio:2007hw, Wang:2015vgv}, we obtain the hard-collinear-scale jet function, while retaining only the matrix elements of soft fields, which correspond to the $\Lambda_b$ LCDA.

\subsubsection{Correlation function at tree level}

\begin{figure}
\centering
\includegraphics[width=0.5 \columnwidth]{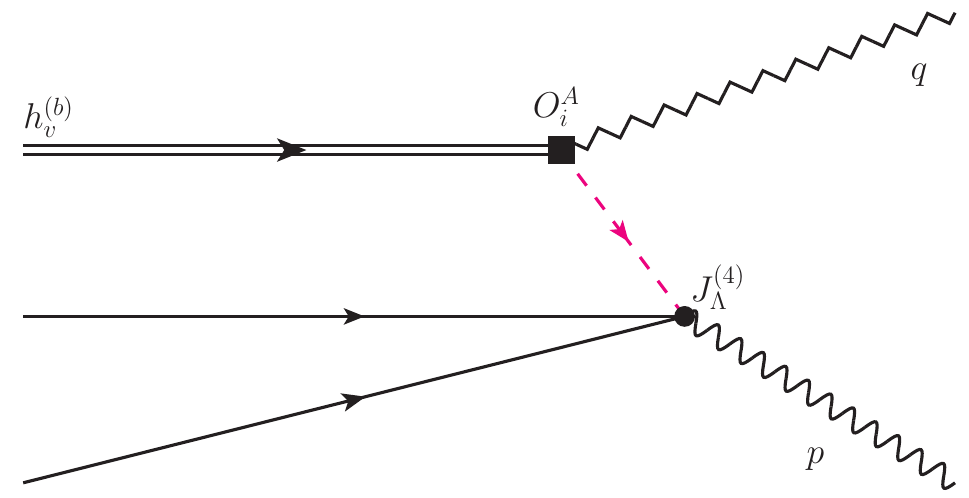}
\vspace*{0.1cm}
\caption{Diagrammatic representation of the correlation function
$\Pi^{\rm A}_{\rm OPE}(\bar n \cdot p)$ at tree level, where the black square denotes the weak transition vertex and the black blob represents
the Dirac structure of the $\Lambda$-baryon current. The double line indicates the $b$-quark in HQET; the strange quark is shown by a pink line, and the soft quarks by solid black lines. In addition, hard-collinear propagators are represented by dashed lines.
.}
\label{fig: tree_correlator}
\end{figure}

At tree level, as shown in Fig.~\ref{fig: tree_correlator}, with $\omega_{1,2} = \bar{n} \cdot k_{1,2}$ corresponding to the light-cone momenta of the up- and down-quarks, the correlation function takes the following form:
\begin{align}
  \Pi^{\rm A, LO}_{\rm OPE}(\bar{n} \cdot p) & \equiv i \int d^4x \, e^{ip \cdot x} \langle 0| T \left[ J_\Lambda^{(4)}(x) \,, O_i^A \right] |\Lambda_b(v,s)\rangle 
\cr 
 &= \int d\omega_1 \, d\omega_2\, F^{(0)}(\omega_1, \omega_2)
\cr 
 &= f_{\Lambda_b}^{(2)} \int_0^\infty \frac{d\omega \,  \tilde\phi_4(\omega) }{\omega - \bar{n} \cdot p - i\epsilon} \
 \frac{\slashed{\bar n}}{2} \, \Gamma_i^A \, u_{\Lambda_b}(v,s) \,,
\end{align} 
where we have defined
\begin{align}
\omega &= \omega_1 + \omega_2 \,,
\\
F^{(0)}(\omega_1, \omega_2) &= f_{\Lambda_b}^{(2)} \, \frac{ \phi_4(\omega_1,\omega_2) }{\omega - \bar{n} \cdot p - i\epsilon} \
 \frac{\slashed{\bar n}}{2} \, \Gamma_i^A \, u_{\Lambda_b}(v,s) \,, 
 \\
 \tilde{\phi}_4(\omega) &= \omega \, \int_0^1  du \,
\phi_4 \left (u \, \omega,  (1 - u) \, \omega \right ) \,. 
\end{align}
The jet function at tree level is simply unity. To obtain this result, we have employed the momentum-space projector for the heavy $\Lambda_b$ baryon, based on the definition of its light-cone distribution amplitudes as derived in Appendix~\ref{App:DAs}. The SCET tree-level result is in exact agreement with the full QCD computation presented in~\cite{Feldmann:2011xf, Wang:2015ndk}.

\subsubsection{Factorization of the correlation function at \texorpdfstring{${\cal O}(\alpha_s)$}{O(alpha\_s)}}

\begin{figure}
\centering
\includegraphics[width=0.8 \columnwidth]{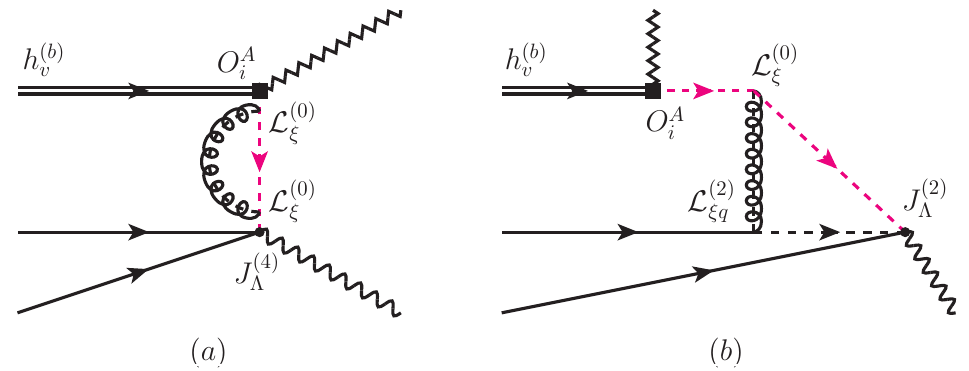}
\vspace*{0.1cm}
\caption{Diagrammatic representation of the non-vanishing one-loop contribution to
$\Pi^{\rm A}_{\rm OPE}(\bar n \cdot p)$ at leading power.}
\label{fig: NLO_correlator}
\end{figure}

We now proceed to determine the NLO contribution to the jet function by expanding the matching condition to ${\cal O}(\alpha_s)$ accuracy.  
To this end, we evaluate the one-loop ${\rm SCET}_{\rm I}$ diagrams using the subleading power SCET Feynman rules collected in Ref.~\cite{Beneke:2018rbh}.

There are only two one-loop diagrams in the light-cone gauge~\cite{DeFazio:2005dx,DeFazio:2007hw}. The first corresponds to the interpolating current $J_{\Lambda}^{(4)}$ combined with the leading-power SCET Lagrangian ${\cal L}_{\xi}^{(0)}$, shown as diagram (a) in Fig.~\ref{fig: NLO_correlator}. The second involves the interpolating current $J_{\Lambda}^{(2)}$ and the power-suppressed SCET Lagrangian ${\cal L}_{\xi q}^{(2)}$, depicted as diagram (b) in Fig.~\ref{fig: NLO_correlator}. Both diagrams contribute at the same power as the tree-level result.

For diagram (a) in Fig.~\ref{fig: NLO_correlator}, the correlation function takes the following form:
\begin{align}
  \Pi^{\rm A, NLO}_{\rm OPE,(a)}(\bar{n} \cdot p) = &  \,\,i \int d^4x  \int d^4y \int d^4z \, e^{ip \cdot x} \langle 0| T \left[ J_\Lambda^{(4)}(x) \,\,  i{\cal L}_{\xi_s}^{(0)}(y) \,\, i{\cal L}_{\xi_s}^{(0)}(z) \,\, O_i^A \right] |\Lambda_b(v,s)\rangle 
\cr 
=&  \int d\omega_1 \, d\omega_2\, F^{(0)}(\omega_1, \omega_2) \,
\frac{\alpha_s \, C_F}{4 \, \pi}  \,
\bigg[ \frac{4}{\epsilon^2} + \frac{1}{\epsilon} \left( 4 \ln \frac{\mu^2}{n \cdot p \, (\omega - \bar n \cdot p)} + 3 \right)
\cr
&+ 2 \ln^2 \frac{\mu^2}{n \cdot p \, (\omega - \bar n \cdot p)} + 3 \ln \frac{\mu^2}{n \cdot p \, (\omega - \bar n \cdot p)}
- \frac{\pi^2}{3} + 7 \bigg] \,,
\end{align} 
where the loop integration is performed in $D = 4 - 2\epsilon$ dimensions. Notably, this diagram shares the same loop structure as the meson case, and we therefore expect the perturbative function to match that of the corresponding diagram in the $B \rightarrow P, V$ transitions. Our calculation confirms that the result for $\Lambda_b$ baryon decay is indeed consistent with the $B$ meson case, as presented in~\cite{DeFazio:2007hw, Gao:2019lta}.

For diagram (b) in Fig.~\ref{fig: NLO_correlator}, we consider the contribution of a hard-collinear gluon attached to either the $d$ or the $u$ quark (assuming isospin symmetry). The correlation function then takes the form:
\begin{align}
  \Pi^{\rm A, NLO}_{\rm OPE,(b)}(\bar{n} \cdot p) = &  \,\,i \int d^4x  \int d^4y \int d^4z \, e^{ip \cdot x} \langle 0| T \left[ J_\Lambda^{(2)}(x) \,\,  i{\cal L}_{\xi q}^{(2)}(y) \,\, i{\cal L}_{\xi_s}^{(0)}(z) \,\, O_i^A \right] |\Lambda_b(v,s)\rangle 
\cr 
=& \, -i\,g_s^2 \, C_F \,\epsilon_{ijk} \int d\omega_1  \int d\omega_2 \int \frac{d^D l}{(2\pi)^D} \int \frac{d^D L}{(2\pi)^D} \times
\cr
& \frac{n \cdot L \,\, n \cdot (p-L) \,\, n \cdot (p-L+l)}
{[L^2 + i 0] [(p-k_1-L)^2 + i 0] [(p-k_1-L+l)^2 + i 0] [l^2 + i 0]} \times \cr
& \langle 0| \, \bar q_u^{\rm T,i}(\omega_1) \, C \, \gamma_5 \, \slashed{n} \, \frac{\slashed{\bar n}}{2} \left[ \frac{\slashed{n}}{2} (\bar{n}^\mu + \frac{\slashed{L}_\perp}{n \cdot L} \gamma_\perp^\mu) + \gamma_\perp^\mu \, {k_2}_{\perp \beta} \, \partial_\perp^\beta \right] \times \cr
& (2 \pi)^4 \delta^4(l + k_2 - L) \, q_d^{\, j}(\omega_2) \, \left[ g_{\mu \nu} - \frac{1}{n \cdot l} (l_\mu \, n_\nu + n_\mu \, l_\nu) \right] \times \cr
& \left[ \bar{n}^\nu - \frac{\slashed{L}_\perp \, \gamma_\perp^\nu}{n \cdot (p - L)} + \frac{\gamma_\perp^\nu \, (\slashed{l}_\perp - \slashed{L}_\perp)}{n \cdot (p - L + l)} \right] 
\, \frac{\slashed{\bar n}}{2} \, \Gamma_i^A \, h^k(0) \, |\Lambda_b(v,s)\rangle \,, \hspace{0.5 cm}
\end{align} 
which can be evaluated using three distinct methods by identifying the transverse derivative
$\partial_\perp^\beta$ acting on the Dirac delta function as $\partial / \partial {k_2}_{\perp \beta}$, $\partial / \partial L_{\perp \beta}$, or $\partial / \partial l_{\perp \beta}$, respectively~\cite{Beneke:2018rbh}. We have explicitly verified that all three approaches yield the same result:
\begin{align}
  \Pi^{\rm A, NLO}_{\rm OPE,(b)}(\bar{n} \cdot p) & =  \int d\omega_1 \, d\omega_2\, F^{(0)}(\omega_1, \omega_2) \,
\frac{\alpha_s \, C_F}{4 \pi}  \,
\bigg\{ -\frac{2}{\epsilon^2} - \frac{1}{\epsilon} \,
\bigg[ 2 \ln \frac{\mu^2}{n \cdot p \, (\omega - \bar{n} \cdot p)} \cr
& + 2 \ln \frac{\omega - \bar{n} \cdot p}{\omega_1 - \bar{n} \cdot p} + \frac{7}{2} \bigg]
 - \ln^2 \frac{\mu^2}{n \cdot p \, (\omega - \bar{n} \cdot p)} - \frac{7}{2} \ln \frac{\mu^2}{n \cdot p \, (\omega - \bar{n} \cdot p)}  
 \cr
& -2 \ln \frac{\omega - \bar{n} \cdot p}{\omega_1 - \bar{n} \cdot p} \ln \frac{\mu^2}{n \cdot p \, (\omega - \bar{n} \cdot p)} - \ln^2 \frac{\omega - \bar{n} \cdot p}{\omega_1 - \bar{n} \cdot p}  \cr 
& + \left( 2 \frac{\omega_1 - \bar{n} \cdot p}{\omega_2} - \frac{3}{2} \right) \ln \frac{\omega - \bar{n} \cdot p}{\omega_1 - \bar{n} \cdot p}
+ \frac{\pi^2}{6} - \frac{15}{2} \bigg\} \,,
\end{align}
where we have employed the Wandzura-Wilczek approximation as given in \refeq{ WW}.

\subsubsection{Leading power contribution}
\label{sec: LP}

The tree-level and one-loop contributions to the A-type correlation function are formally suppressed by $\lambda^2$ compared to the leading-power contribution, which begins at the two-loop level.
It is justified to neglect the leading-power contribution in our analysis. This conclusion is consistent with~\cite{Wang:2011uv}, which observes that the numerical value of the leading-power form factor is approximately one order of magnitude smaller than that governed by soft processes.

\begin{figure}
\centering
\hspace{0.2cm}
    \begin{subfigure}[b]{0.48\textwidth}
\includegraphics[width=0.9 \textwidth]{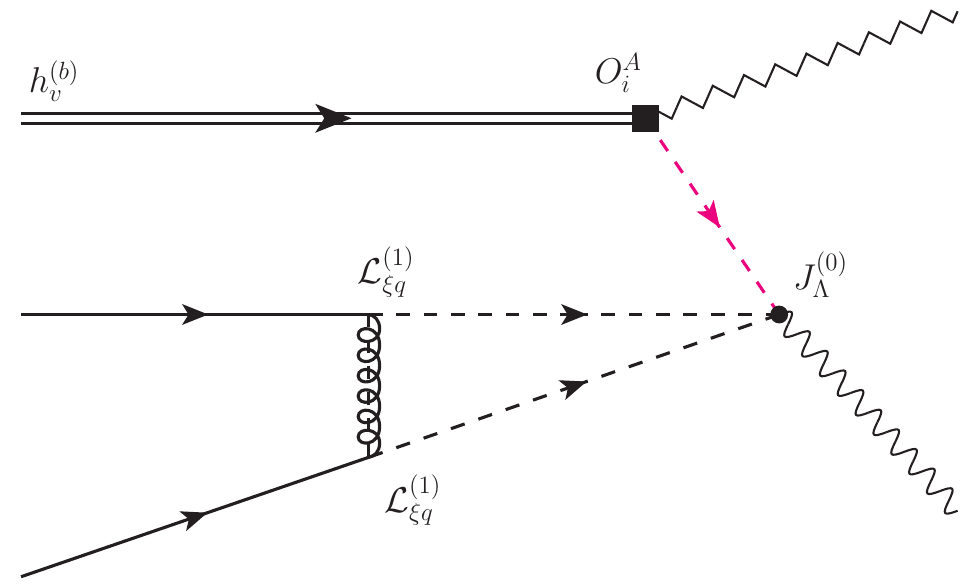}
\caption{}
\label{fig:scaleless_correlator}
    \end{subfigure}
    \hfill
    \begin{subfigure}[b]{0.48\textwidth}
\includegraphics[width=0.9 \textwidth]{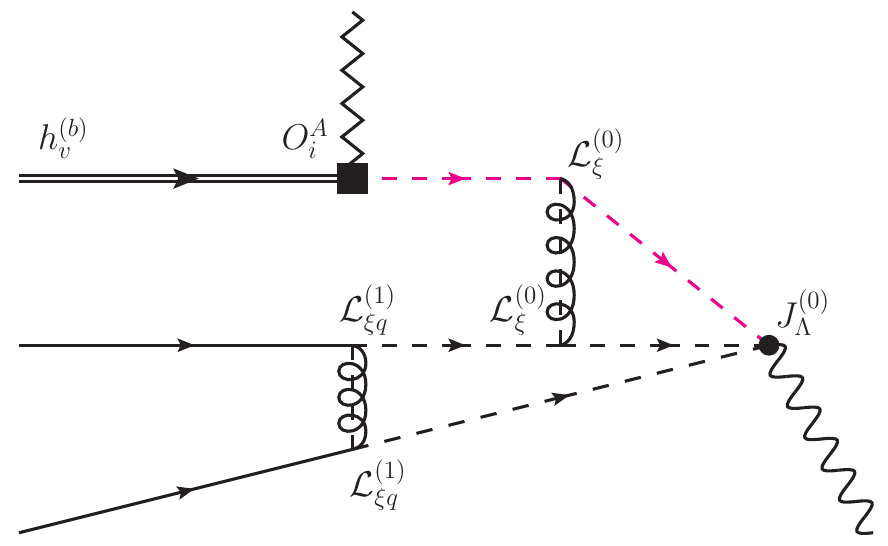}
\caption{}
	\label{fig:twoloop_correlator}
    \end{subfigure}
    \caption{Diagram (a) shows the leading-power one-loop representation of $\Pi^{\rm A}_{\rm OPE}(\bar n \cdot p)$ , while diagram (b) illustrates a possible leading-power two-loop contribution to $\Pi^{\rm A}_{\rm OPE}(\bar n \cdot p)$.}
\end{figure}
For the leading-power contribution, the interpolating current is $J_{\Lambda}^{(0)}$. At one-loop level, there are two soft-collinear interactions ${\cal L}_{\xi q}^{(1)}$ that convert the $u$ and $d$ quarks in $J_{\Lambda}^{(0)}$ from hard-collinear to soft via the exchange of a hard-collinear gluon, as illustrated in Fig.~\ref{fig:scaleless_correlator},
\begin{align}
 \Pi^{\rm A}_{0} \equiv i \int d^4x  \int d^4y \int d^4z \, e^{ip \cdot x} \langle 0| T \left[ i{\cal L}_{\xi_u q_u}^{(1)}(y) \,\, i{\cal L}_{\xi_d q_d}^{(1)}(z) \,\, J_\Lambda^{(0)}(x) \,\, O_i^A \right] |\Lambda_b(v,s)\rangle \,.
\end{align}
However, the loop integral of this diagram is scaleless, in agreement with Ref.~\cite{Wang:2015ndk}.
Consequently, the leading-power contribution to the A-type correlation function arises at the two-loop level. A schematic diagram in light-cone gauge is shown in Fig.~\ref{fig:twoloop_correlator}. 

\subsubsection{Factorization-scale independence}

The factorization formula for the A-type correlation function is
\begin{equation}
\begin{aligned}
  \Pi^{\rm A}_{\rm OPE}(\bar{n} \cdot p) & \equiv \Pi^{\rm A, LO}_{\rm OPE}(\bar{n} \cdot p) + \Pi^{\rm A, NLO}_{\rm OPE,(a)}(\bar{n} \cdot p) + \Pi^{\rm A, NLO}_{\rm OPE,(b)}(\bar{n} \cdot p)  + {\cal O}(\alpha_s^2)
\\[2pt] 
 &= f_{\Lambda_b}^{(2)}(\mu) \int d\omega_1 \, d\omega_2\,  \frac{ \phi_4(\omega_1,\omega_2,\mu) }{\omega - \bar{n} \cdot p -i\epsilon} \, J\left ( \frac{\mu^2}{\bar n \cdot p \, \omega_i}, \frac{\omega_i}{\bar n \cdot p} \right ) \, 
 \frac{\slashed{\bar n}}{2} \, \Gamma_i^A \, u_{\Lambda_b}(v,s) + {\cal O}(\alpha_s^2)
 \, ,
 \label{eq: NLO correlator}
\end{aligned}
\end{equation}
where the jet function is
\begin{align}
  J\left ( \frac{\mu^2}{\bar n \cdot p \, \omega_i},
\frac{\omega_i}{\bar n \cdot p} \right )  = & \,\, 1 +
\frac{\alpha_s(\mu)}{4 \, \pi} \, C_F \,
\bigg \{ 
\ln^2 \frac{\mu^2}{n \cdot p \, (\omega - \bar n \cdot p)} - \, \frac{1}{2} \, \ln \frac{\mu^2}{n \cdot p \, (\omega - \bar n \cdot p)}  
 \cr
& -2 \,  \ln \frac{\omega - \bar n \cdot p}{\omega_1 - \bar n \cdot p} \, \ln \frac{\mu^2}{n \cdot p \, (\omega - \bar n \cdot p)} - \, \ln^2 \frac{\omega - \bar n \cdot p}{\omega_1 - \bar n \cdot p}  \cr 
& + \left( 2\, \frac{\omega_1 - \bar n \cdot p}{\omega_2} -  \frac{3}{2} \right) \, \ln \frac{\omega - \bar n \cdot p}{\omega_1 - \bar n \cdot p}
- \frac{\pi^2}{6} - \frac{1}{2}  \, \bigg \}  \,.
\end{align}
This result agrees with~\cite{Wang:2015ndk}, which calculates the $\Lambda_b \rightarrow \Lambda$ form factors using the method of regions~\cite{Beneke:1997zp,Jantzen:2011nz}.


Having obtained the one-loop factorization formulae, we now explicitly verify the factorization scale independence of the correlation functions $C_i^A(n\cdot p,\mu) \, \Pi^{\rm A}_{\rm OPE}(\bar{n} \cdot p)$ at one loop. The RG equation for the hard function reads
\begin{align}
 \frac{d}{d \ln \mu} \, C_i^A(n\cdot p,\mu)  = \gamma^A \, C_i^A(n\cdot p,\mu)\,,
 \label{eq:RGE of hard functions}
\end{align}
where~\cite{Bauer:2000yr}
\begin{align}
 \gamma^A = -\Gamma_{\rm cusp}(\alpha_s) \, \ln {\frac{\mu}{n \cdot p}} + \tilde{\gamma}(\alpha_s) = {\frac{\alpha_s(\mu)}{4\pi}} \, C_F  \left(-4 \ln {\frac{\mu}{n \cdot p}} - 5 \right) + {\cal O}(\alpha_s^2) \,.
\end{align}
For the jet function, we find
\begin{align}
 \frac{d}{d \ln \mu} \, J\left ( \frac{\mu^2}{\bar{n} \cdot p \, \omega_i},  
\frac{\omega_i}{\bar{n} \cdot p} \right )  =  \frac{\alpha_s(\mu)}{4\pi} \, C_F \,  
 \left [ 4 \ln \frac{\mu^2}{n \cdot p \, (\omega - \bar{n} \cdot p)}  
- 4 \ln \frac{\omega - \bar{n} \cdot p}{\omega_1 - \bar{n} \cdot p} - 1  \right ]\,.  
\end{align}
We also obtain the one-loop evolution equation for the $\Lambda_b$-baryon DA $\phi_4(\omega_1,\omega_2,\mu)$, which agrees with~\cite{Wang:2015ndk}:
\begin{eqnarray}
&& \hspace{0.5cm}
\int_0^{\infty} d\omega_1 \int_0^{\infty} d\omega_2 \,
\frac{1}{\omega_1 + \omega_2 - \bar{n} \cdot p - i 0} \,\,
\frac{d}{d \ln \mu} \left[ f_{\Lambda_b}^{(2)}(\mu) \, \phi_4(\omega_1,\omega_2,\mu) \right] \nonumber \\
&& = - \frac{\alpha_s(\mu)}{4\pi} \, C_F \,
\int_0^{\infty} d\omega_1 \int_0^{\infty} d\omega_2 \,
\frac{1}{\omega_1 + \omega_2 - \bar{n} \cdot p - i 0} \, \nonumber \\
&& \hspace{0.5cm} \times \left[ 4 \ln \frac{\mu}{\omega - \bar{n} \cdot p}
- 4 \ln \frac{\omega - \bar{n} \cdot p}{\omega_1 - \bar{n} \cdot p} - 5 \right]
\left[ f_{\Lambda_b}^{(2)}(\mu) \, \phi_4(\omega_1,\omega_2,\mu) \right] \,.
\end{eqnarray}
From the above results, we can readily deduce
\begin{eqnarray}
\frac{d}{d \ln \mu} \left[ C_i^A(n\cdot p,\mu) \, \Pi^{\rm A}_{\rm OPE}(\bar{n} \cdot p) \right]
= - \frac{\alpha_s(\mu)}{4\pi} \, C_F \, f_{\Lambda_b}^{(2)}(\mu) 
\int_0^{\infty} d\omega_1 \int_0^{\infty} d\omega_2 \,
\frac{\phi_4(\omega_1,\omega_2,\mu)}{\omega - \bar{n} \cdot p - i 0} \,.
\label{eq:scale dependence of the correlator}
\end{eqnarray}

The residual $\mu$-dependence in \refeq{scale dependence of the correlator} stems from the ultraviolet renormalization of the baryon current:
\begin{eqnarray}
\frac{d}{d \ln \mu} \ln f_{\Lambda}(\mu) = - \left( \frac{\alpha_s(\mu)}{4\pi} \right)^k \,
\gamma_{\Lambda}^{(k)} \,,
\label{scale dependence of the baryonic current}
\end{eqnarray}
with $\gamma_{\Lambda}^{(1)} = 4/3$~\cite{Chernyak:1987nu,Braun:2000kw}.
By distinguishing the renormalization scales for the interpolating current of the $\Lambda$-baryon and the weak transition current in QCD from the factorization scale (see the next section for details), we arrive at the conclusion that the factorization-scale dependence cancels completely in the factorized expressions of the correlation functions $C_i^A(n\cdot p,\mu) \, \Pi^{\rm A}_{\rm OPE}(\bar{n} \cdot p)$ at one loop.

\subsubsection{LCSR of \texorpdfstring{$\xi_\Lambda(n \cdot p)$}{xi\_Lambda(n·p)} at \texorpdfstring{${\cal O}(\alpha_s)$}{O(alpha\_s)}}
\label{section: resummation improved LCSR}

The dispersion forms of the correlation functions are derived using spectral representations in Appendix A of Ref.~\cite{Wang:2015ndk}. By comparing the partonic and hadronic representations of the correlation functions, subtracting the continuum-state contributions, and performing a Borel transformation, we obtain the sum rule for the form factor:
\begin{align}
n \cdot p \, f_\Lambda(\nu) \, \exp\left(-\frac{m_\Lambda^2}{n \cdot p \, \omega_M}\right) \, \xi_\Lambda(n \cdot p) 
&= f_{\Lambda_b}^{(2)} \int_0^{\omega_s} d\omega' \, \phi_{\rm 4, eff}(\omega', \mu, \nu) \, e^{-\omega'/\omega_M} \,,
\label{eq:NLOexact}
\end{align}
where $\nu$ denotes the renormalization scale of the baryon current. It is evident that the $\ln \nu$ dependence must be separated from the jet function.

The effective ``distribution amplitude'' $\phi_{4, \rm eff}(\omega^{\prime},\mu, \nu)$ is given by
\begin{eqnarray}
\phi_{4, \rm eff}(\omega^{\prime},\mu, \nu) &=& \tilde{\phi}_{4}(\omega^{\prime},\mu)
+ \frac{\alpha_s(\mu)}{4 \, \pi} \, C_F \,
\bigg\{ \int_0^{\omega^{\prime}} d \omega \,
\left[ \frac{2}{\omega - \omega^{\prime}} \,
\ln \frac{\mu^2}{n \cdot p \, (\omega^{\prime} - \omega)} \right]_{\oplus}
\, \tilde{\phi}_{4}(\omega,\mu) \nonumber
\\
&& - \, 2 \, \omega^{\prime} \int_0^{\omega^{\prime}} d \omega \,
\left[ \frac{1}{\omega - \omega^{\prime}} \,
\ln \frac{\omega^{\prime} - \omega}{\omega^{\prime}} \right]_{\oplus} \, \phi_4(\omega,\mu) \nonumber
\\
&& - \, \omega^{\prime} \int_{\omega^{\prime}}^{\infty} d \omega \,
\bigg[ \frac{\omega}{\omega^{\prime}} \, \ln^2 \frac{\mu^2}{n \cdot p \, \omega^{\prime}}
- 2 \, \ln \frac{\mu^2}{n \cdot p \, \omega^{\prime}} \,
\ln \frac{\omega - \omega^{\prime}}{\omega^{\prime}}
- \frac{11}{2} \, \ln \frac{\omega - \omega^{\prime}}{\omega^{\prime}} \nonumber
\\
&& \hspace{2.6cm} - \, \frac{\pi^2 + 1}{2} \, \frac{\omega}{\omega^{\prime}}
+ \left( \frac{2 \pi^2}{3} - \frac{11}{2} \right) \bigg] \,
\frac{d \phi_4(\omega,\mu)}{d \omega} \nonumber
\\
&& - \int_{\omega^{\prime}}^{\infty} d \omega \,
\left[ \ln^2 \frac{\mu^2}{n \cdot p \, \omega^{\prime}}
+ 2 \, \ln \frac{\omega - \omega^{\prime}}{\omega}
- \frac{\pi^2 + 1}{2} \right] \phi_4(\omega,\mu) \nonumber
\\
&& - \int_0^{\omega^{\prime}} d \omega \,
\left[ 2 \, \ln \frac{\mu^2}{n \cdot p \, (\omega^{\prime} - \omega)}
+ \frac{1}{2} \, \ln \frac{\nu^2}{n \cdot p \, (\omega^{\prime} - \omega)} \right]
\frac{d \tilde{\phi}_4(\omega,\mu)}{d \omega} \bigg\} \,, \hspace{0.5cm}
\label{eq:phi4eff}
\end{eqnarray}
where the $\Lambda_b$-baryon DA $\phi_4(\omega_1,\omega_2,\mu)$ is abbreviated as $\phi_4(\omega,\mu)$ for brevity, since
\[
\phi_4(\omega_1, \omega_2, \mu) = \phi_4(u \, \omega, (1 - u) \, \omega, \mu)
\]
is assumed to be independent of the momentum fraction $u$, as motivated by Refs.~\cite{Feldmann:2011xf,Bell:2013tfa,Ball:2008fw}.
Within this approximation, $\tilde{\phi}_4(\omega,\mu)$ is given by
\[
\tilde{\phi}_4(\omega, \mu) = \omega \int_0^1 du \,
\phi_4(u \, \omega, (1 - u) \, \omega, \mu) = \omega \, \phi_4(\omega,\mu) \,.
\]
The $\oplus$ function is defined as
\begin{eqnarray}
\int_0^{\infty} d \omega \, \left[ f(\omega,\omega^{\prime}) \right]_{\oplus} \, g(\omega)
= \int_0^{\infty} d \omega \, f(\omega,\omega^{\prime}) \,
\left[ g(\omega) - g(\omega^{\prime}) \right] \,.
\end{eqnarray}
From \refeq{NLOexact}, we can derive the scaling law for the ``soft'' form factor:
\begin{equation}
\xi_\Lambda(n \cdot p) \sim \lambda^6 \,.
\end{equation}
It is suppressed by $\lambda^2$ compared to the leading-power form factors,
which is consistent with the results in Ref.~\cite{Wang:2011uv}.

\subsection{LCSR for B-type form factor}

\begin{figure}[tpb]
\centering 
 \includegraphics[width=0.45 \columnwidth]{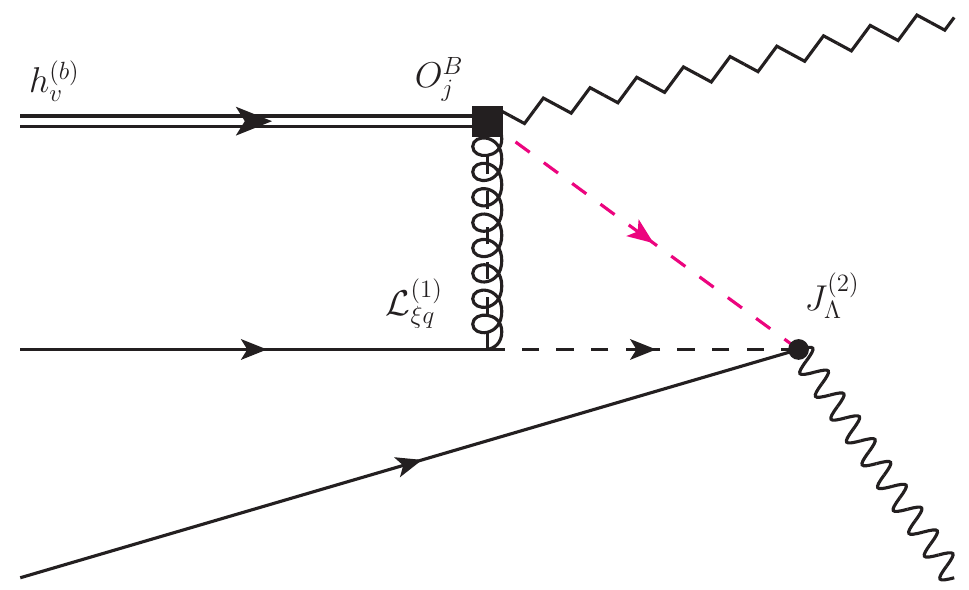}

\caption{Diagrammatic representation of the leading-power
$\Pi^{\rm B}_{\rm OPE}(u, \bar n \cdot p)$ at leading order.}\label{fig: B-type OPE} 
\end{figure}

We now define the correlation function between the B-type weak decay current and
the interpolating current $J_\Lambda$ to extract the non-local form factor $\Delta\xi_\Lambda^B(u,n \cdot p)$, which corresponds to the hard-scattering corrections discussed in Ref.~\cite{Feldmann:2011xf}.
Accordingly, we define
\begin{align}
\Pi^{\rm B}(u, \bar{n} \cdot p) &\equiv i \, n \cdot p \int d^4x \, e^{i p \cdot x} \int \frac{da}{2\pi} \, e^{-i u \, n \cdot p \, a} \,
\langle 0 | T \left[ J_\Lambda(x) \,, O_j^B(a) \right] |\Lambda_b(v,s)\rangle \,,
\end{align}
with $O_j^B(a) = \bar{\xi}_{s}(0) \, \Gamma_j^B \, A_{hc\perp}^{\nu}(a n) \, h(0)$.

Due to the isospin symmetry of the strong interaction, there is only one contributing diagram at ${\cal O}(\alpha_s)$, as shown in Fig.~\ref{fig: B-type OPE}. By exchanging a hard-collinear gluon, the relevant ingredients are the interpolating current $J_{\Lambda}^{(2)}$ and the subleading-power SCET Lagrangian ${\cal L}_{\xi q}^{(1)}$. Owing to the additional ${A}_{hc}^{\perp}$ field in $O_j^B(a)$, this correlation function is of the same power as the A-type correlation functions, and takes the following form:
\begin{align}
\Pi^{\rm B,LO}_{\rm OPE}(u, \bar{n} \cdot p) =\, & i \, n \cdot p \int d^4x \, e^{i p \cdot x} \int d^4y \int \frac{da}{2\pi} \, e^{-i u \, n \cdot p \, a} \cr 
& \times \langle 0| T \left[ J_\Lambda^{(2)}(x) \,, i{\cal L}_{\xi q}^{(1)}(y) \,, O_j^B(a) \right] |\Lambda_b(v,s)\rangle \,.
\end{align}

This correlation function vanishes due to rotational invariance in the transverse plane (which implies an odd number of $\gamma_{\perp}$ matrices in the trace over spinor space). This is why the insertion of ${\cal L}_{\xi q}^{(1)}$ yields a vanishing contribution in diagram (b) of Fig.~\ref{fig: NLO_correlator} \cite{DeFazio:2005dx}. In Ref.~\cite{Feldmann:2011xf}, this diagram is evaluated using the full momentum in the numerator of the hard-collinear quark propagators, while employing the light-cone projector to preserve rotational invariance in the transverse plane. According to their procedure, only the subleading transverse momentum components in the numerators of the hard-collinear $s$-quark and $u$ ($d$)-quark propagators contribute. This implies that their $\Delta\xi_\Lambda$ is formally $\lambda^2$ ($\alpha_s$)-suppressed relative to the soft form factor $\xi_\Lambda$. This power counting is consistent with their numerical result $\Delta \xi_\Lambda/\xi_\Lambda \simeq -0.8\%$.

\subsection{LCSR for C-type non-factorisable contributions}
\label{Sec: NF contributions}

The C-type annihilation operator corresponds to the non-factorizable contribution \cite{Feldmann:2023plv,Wang:2015ndk,Wang:2014jya}. For convenience, we define new operators $O^C_\pm(a)$ with explicit colour indices as:
\begin{equation}
O^C_\pm (a) = \bar{\xi}_{s}^{k_1}(0)(1+\gamma_5)\gamma_\perp^\mu
\, h^{k_2}(0)~ \bar{\xi}_{q}^{l_1}(an)\gamma^\perp_\mu(1 \pm \gamma_5)
\chi_q^{l_2}(0)~. 
\end{equation}
The C-type $\gamma^\ast$-to-$\Lambda_b$-baryon correlation function is defined as
\begin{align}
\Pi^{\rm C,i}(u, \bar{n} \cdot p) & \equiv i \, n \cdot p \int d^4x \, e^{ip \cdot x} \int \frac{da}{2 \pi} \, e^{-i u \, n \cdot p \, a} \langle \gamma^\ast(q,\lambda)| T \left[ J_\Lambda(x) \,, Q_i^C(0,a,0) \right] |\Lambda_b(v,s)\rangle \,, \nonumber \\
\Pi^{\rm C,\pm}(u, \bar{n} \cdot p) & \equiv i \, n \cdot p \int d^4x \, e^{ip \cdot x} \int \frac{da}{2 \pi} \, e^{-i u \, n \cdot p \, a} \langle \gamma^\ast(q,\lambda)| T \left[ J_\Lambda(x) \,, O_{\pm}^C(a) \right] |\Lambda_b(v,s)\rangle \,.
\end{align}
Then we have the following relations:
\begin{equation}
\begin{aligned}
\Pi^{\rm C,1}(u, \bar{n} \cdot p) &= \delta_{k_1 k_2}\delta_{l_1 l_2}\, \Pi^{\rm C,-}(u, \bar{n} \cdot p)~, \quad 
\Pi^{\rm C,2}(u, \bar{n} \cdot p) = \delta_{k_1 l_2}\delta_{l_1 k_2}\, \Pi^{\rm C,-}(u, \bar{n} \cdot p)~, \\
\Pi^{\rm C,3}(u, \bar{n} \cdot p) &= \delta_{k_1 k_2}\delta_{l_1 l_2}\, \Pi^{\rm C,+}(u, \bar{n} \cdot p)~, \quad 
\Pi^{\rm C,4}(u, \bar{n} \cdot p) = \delta_{k_1 l_2}\delta_{l_1 k_2}\, \Pi^{\rm C,+}(u, \bar{n} \cdot p)~.
\end{aligned}
\label{eq: CiCpmrelation}
\end{equation}

On the hadronic side, the contribution of the $\Lambda$ baryon to the correlation function
is given by
\begin{align}
 \Pi^{\rm C,i} \big|_{\rm res.} & \simeq n \cdot p \int \! \frac{da}{2\pi}~e^{-iu\, n \cdot p\, a} \sum_{s'} 
   \frac{ 
\langle 0| J_\Lambda| \Lambda(p,s')\rangle \, \langle \Lambda(p,s')\, \gamma^\ast(q,\lambda)|\,{Q}_{i}^C(0,a,0) \,| \Lambda_b(v,s) \rangle }{m_\Lambda^2- p^2}
\cr 
&= g_{\rm em} \, M_{\Lambda_b}^2 \, \bar{\epsilon}_\nu^\ast(q,\lambda)\,\, \Delta_i\xi_\Lambda^C(u,n \cdot p) \, \frac{(n \cdot p) \, f_\Lambda (\nu)}{m_\Lambda^2/(n \cdot p)- (\bar{n} \cdot p)} \, \frac{\slashed {\bar n}}{2} \, (1+\gamma_5)\, \gamma_{\perp}^{\nu} \,  u_{\Lambda_b}(v,s)  \,.
\end{align}

\begin{figure}
\begin{center}
\includegraphics[width=0.5 \columnwidth]{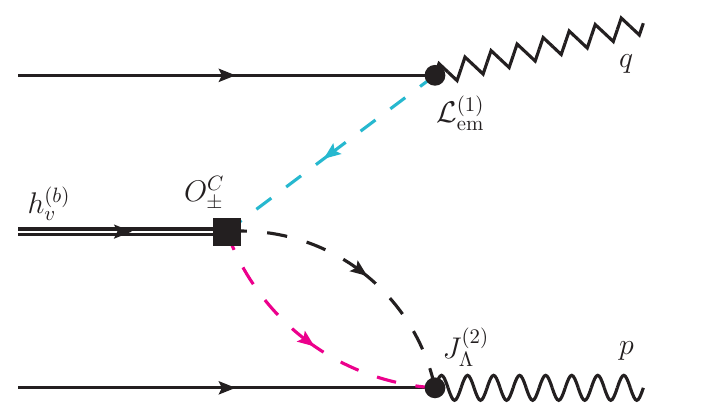}
\end{center}
\caption{Diagrammatic representation of the leading-power
$\Pi^{\rm C, \pm}_{\rm OPE}(u, \bar n \cdot p)$ at leading order. Blue dashed line indicate anti-hard-collinear propagator. }
\label{fig: NF}
\end{figure}

On the partonic side, the OPE of $\Pi^{\rm C,i}(u, \bar{n} \cdot p)$ is obtained via the relation in \refeq{ CiCpmrelation}. By applying isospin symmetry of the strong interaction, only one diagram appears at leading order, as shown in Fig.~\ref{fig: NF}, where the photon is emitted from a spectator quark. The relevant components are the interpolating current $J_{\Lambda}^{(2)}$ and the subleading-power SCET electromagnetic Lagrangian:
\begin{equation}
   {\cal L}_{\rm em}^{(1)} = g_{\rm em} \sum_q Q_q\, \bar{\chi}_q \,\slashed{A}_{\overline{hc} \perp}^{(\rm em)} \, q_q\,,
\end{equation}
where $Q_q$ is the electric charge of the quark $q$. Inserting the effective electromagnetic interaction vertex yields the following expression for Fig.~\ref{fig: NF}:
\begin{equation}
\begin{aligned}
\Pi^{\rm C,\pm}_{\rm OPE}(u, \bar{n} \cdot p) =&\, i \, n \cdot p \int d^4x \, e^{ip \cdot x} \int d^4y \int \frac{da}{2 \pi} \, e^{-i u \, n \cdot p \, a} \times \\
& \hspace{3.5cm} \langle \gamma^\ast(q,\lambda)| T \left[ J_\Lambda^{(2)}(x)\,,\, O_\pm^C(a)\,,\, i{\cal L}_{\rm em}^{(1)}(y) \right] |\Lambda_b(v,s)\rangle \\
=&\, i\, g_{\rm em} \, Q_u \, \bar{\epsilon}_\nu^\ast(q,\lambda)\, \epsilon_{i l_1 k_1} \int d\bar{\omega}_1 \, \frac{1}{n \cdot q - \bar{\omega}_1} \int d\omega_2 \times \\
& \int \frac{d^D l}{(2\pi)^D} \, \delta\left(u - \frac{n \cdot l}{n \cdot p}\right) \frac{n \cdot l}{l^2 + i0} \, \frac{n \cdot (p - l)}{(p - l)^2 - n \cdot (p - l) \, \omega_2 + i0} \times \\
& \hspace{-1cm} \langle 0| \left[\bar{q}_d^{\mathrm{T},i}(\omega_2) \, C \, \gamma_5 \, \slashed{n} \, \frac{\slashed{\bar{n}}}{2} \, \gamma_\perp^\mu \, (1 \pm \gamma_5) \, \frac{\slashed{n}}{2} \, \gamma_\perp^\nu \, q_u^{l_2}(\bar{\omega}_1)\right] \, \frac{\slashed{\bar{n}}}{2} \, \gamma^\perp_\mu \, (1 - \gamma_5)\, h^{k_2}(0) \, |\Lambda_b(v,s)\rangle\,,
\end{aligned}
\end{equation}
where $\bar{\omega}_1 = n \cdot k_1$ and $\omega_2 = \bar{n} \cdot k_2$. For clarity, we have focused on the scenario where the photon is emitted from the $u$ quark.
Owing to isospin symmetry, the results for photon emission from the $d$ quark are identical, with the only modification being the substitution $Q_u \to Q_d$, as we disregard the masses of the light quarks \cite{Feldmann:2023plv}.

Since the LCSRs are derived using a dispersion relation of the form
\begin{align}
    \Pi^{\rm C,\pm}_{\rm OPE}(u, \bar{n} \cdot p) =
    \frac{1}{\pi} \, \int_0^\infty d\omega \, \frac{\Im \Pi^{\rm C,\pm}_{\rm OPE}(u,\omega)}{\omega - \bar{n} \cdot p - i\epsilon} \,,
\end{align}
we only require the imaginary part of the correlator. The loop integration yields
\begin{align}
    &\quad i\, \mu^{4-D} \int \frac{d^D l}{(2\pi)^D} \, \delta\left(u - \frac{n \cdot l}{n \cdot p}\right) 
    \frac{n \cdot l}{l^2 + i0} \, \frac{n \cdot (p - l)}{(p - l)^2 - n \cdot (p - l) \omega_2 + i0} \cr
    =& \,- \frac{1}{16\pi^2}\, u \bar{u} \, \theta(u) \theta(\bar{u})\, (n \cdot p)^2 
    \left[ \frac{1}{\epsilon} + \ln\left( \frac{-\mu^2}{n \cdot p \, (\bar{n} \cdot p - \omega_2)} \right) 
    - \ln(u \bar{u}) \right]\,.
\end{align}
Then, we obtain the imaginary part of the correlation function:
\begin{align}
\Im \Pi^{\rm C,\pm}_{\rm OPE}(u,\omega) =& \,\frac{1}{16\pi}\, g_{\rm em} \, Q_u \, 
\bar{\epsilon}_\nu^\ast(q,\lambda)\, u \bar{u} \, \theta(u) \theta(\bar{u})\, (n \cdot p)^2 
\int d\bar{\omega}_1 \int d\omega_2 \, \frac{\theta(\omega - \omega_2)}{n \cdot q - \bar{\omega}_1} \times \cr
& \hspace{-1.8cm} \epsilon_{i l_1 k_1} 
\langle 0| \left[\bar{q}_u^{\mathrm{T},l_2}(\bar{\omega}_1) \, C \, \gamma_\perp^\nu \, \slashed{n} \, (1 \pm \gamma_5)\, 
\gamma_\perp^\mu \, \gamma_5 \, q_d^i(\omega_2) \right] 
\frac{\slashed{\bar{n}}}{2} \, \gamma^\perp_\mu \, (1 - \gamma_5)\, h^{k_2}(0) \, |\Lambda_b(v,s)\rangle\,.
\end{align}

The above expression depends on two opposite light-cone projections of the light-quark momenta in the $\Lambda_b$ baryon, requiring the momentum-space projector $M^{(2)}(\bar{\omega}_1,\omega_2)$ defined in \refeq{newlcproj2}. We then obtain:
\begin{align}
\Im \Pi^{\rm C,1}_{\rm OPE}(u,\omega) =& \,-\frac{1}{8\pi}\, g_{\rm em} \, Q_u \, f^{(2)}_{\Lambda_b} \, 
\bar{\epsilon}_\nu^\ast(q,\lambda)\, u \bar{u} \, \theta(u) \theta(\bar{u})\, (n \cdot p)^2 
\int d\bar{\omega}_1 \int d\omega_2 \, \frac{\theta(\omega - \omega_2)}{n \cdot q - \bar{\omega}_1} \times \cr
& \hspace{-1.0cm} \left(\chi_2(\bar{\omega}_1,\omega_2) + \chi_{42}^{(ii)}(\bar{\omega}_1,\omega_2)\right) 
\left(\frac{g_\perp^{\nu \mu}}{2} - \frac{i \epsilon^{\perp^\nu \perp^\mu \{\bar{n}\} \{n\}}}{4} \right) 
\frac{\slashed{\bar{n}}}{2} \, \gamma^\perp_\mu \, (1 - \gamma_5)\, u_{\Lambda_b}(v,s)\,, \cr
\Im \Pi^{\rm C,3}_{\rm OPE}(u,\omega) =& \,-\frac{1}{8\pi}\, g_{\rm em} \, Q_u \, f^{(2)}_{\Lambda_b} \, 
\bar{\epsilon}_\nu^\ast(q,\lambda)\, u \bar{u} \, \theta(u) \theta(\bar{u})\, (n \cdot p)^2 
\int d\bar{\omega}_1 \int d\omega_2 \, \frac{\theta(\omega - \omega_2)}{n \cdot q - \bar{\omega}_1} \times \cr
& \hspace{-1.0cm} \left(\chi_2(\bar{\omega}_1,\omega_2) + \chi_{42}^{(ii)}(\bar{\omega}_1,\omega_2)\right) 
\left(\frac{g_\perp^{\nu \mu}}{2} + \frac{i \epsilon^{\perp^\nu \perp^\mu \{\bar{n}\} \{n\}}}{4} \right) 
\frac{\slashed{\bar{n}}}{2} \, \gamma^\perp_\mu \, (1 - \gamma_5)\, u_{\Lambda_b}(v,s)\,.
\end{align}
The color structure leads to the relations:
\begin{equation}
    \Im \Pi^{\rm C,1}_{\rm OPE}(u,\omega) = -\Im \Pi^{\rm C,2}_{\rm OPE}(u,\omega)\,, \quad 
    \Im \Pi^{\rm C,3}_{\rm OPE}(u,\omega) = -\Im \Pi^{\rm C,4}_{\rm OPE}(u,\omega)\,.
\end{equation}
In our convention, $\gamma_5 = i \gamma^0 \gamma^1 \gamma^2 \gamma^3$. For the Levi-Civita symbol, we use $\epsilon^{0123} = 1$, and adopt the shorthand notation 
$\epsilon^{\mu\nu\rho\sigma} q_\sigma \equiv \epsilon^{\mu\nu\rho\{q\}}$.
In $D = 4$ dimensions, the following identities hold:
\begin{align}
    \left( \frac{g_\perp^{\nu \mu}}{2} - \frac{i \epsilon^{\perp^\nu \perp^\mu \{\bar{n}\} \{n\}}}{4} \right) 
    \frac{\slashed{\bar{n}}}{2} \, \gamma^\perp_\mu \, (1 - \gamma_5) &= 0\,, \cr
    \left( \frac{g_\perp^{\nu \mu}}{2} + \frac{i \epsilon^{\perp^\nu \perp^\mu \{\bar{n}\} \{n\}}}{4} \right) 
    \frac{\slashed{\bar{n}}}{2} \, \gamma^\perp_\mu \, (1 - \gamma_5) &= 
    \frac{\slashed{\bar{n}}}{2} \, \gamma_\perp^\nu \, (1 - \gamma_5)\,,
\end{align}
which lead to the final results:
\begin{align}
    \Delta_1 \xi_\Lambda^C(u, n \cdot p) &= \Delta_2 \xi_\Lambda^C(u, n \cdot p) = 0\,, \cr
    \Delta_3 \xi_\Lambda^C(u, n \cdot p) &= -\Delta_4 \xi_\Lambda^C(u, n \cdot p) 
    = -\Delta \xi_\Lambda^C(u, n \cdot p)\,.
\end{align}

The sum rules for the C-type form factors are
\begin{align}
\Delta\xi_\Lambda^C(u,n \cdot p) =& \, \frac{f^{(2)}_{\Lambda_b}}{f_\Lambda(\nu)}\, \frac{Q_u+Q_d}{8\pi^2}\, \frac{n\cdot p}{M^2_{\Lambda_b}}\, u \bar u \, \theta(u) \theta(\bar u) \,\, \times 
\cr
& \hspace{-1.5cm} \int_0^{\omega_s} d\omega \int_0^\omega d\omega_2 \int_0^\infty d\bar \omega_1 \, \frac{\chi_2(\bar\omega_1,\omega_2) + \chi_{42}^{(ii)}(\bar\omega_1,\omega_2)}{n\cdot q-\bar \omega_1+i\epsilon}\, {\rm exp}\left(\frac{m_\Lambda^2 - n \cdot p \, \omega}{n \cdot p \, \omega_M}\right)\,.
\label{eq:Ctypesumrule}
\end{align}
Our result agrees with Ref.\cite{Feldmann:2023plv} up to a sign difference.\footnote{A minus sign appears to be missing in Eq.(3.32) of Ref.\cite{Feldmann:2023plv}.} From \refeq{Ctypesumrule}, we can derive the scaling law for C-type form factors:
\begin{equation}
    \Delta_3\xi_\Lambda^C(u,n \cdot p) \sim \Delta_4\xi_\Lambda^C(u,n \cdot p) \sim \lambda^6\,,
\end{equation}
while $\Delta_1\xi_\Lambda^C(u,n \cdot p) = \Delta_2\xi_\Lambda^C(u,n \cdot p) = 0$ at $\mathcal{O}(\lambda^6)$. From the above analysis, the relation in \refeq{deltaCA} reduces to:
\begin{align}
    &\Delta C_1^A(q^2) = 16 \, \pi^2 \, \frac{\alpha_{\rm e}}{4\pi} \, \frac{1}{q^2} \, M_{\Lambda_b}^2 \int \! du \left[C_3^C-C_4^C\right](n \cdot p,\bar{u}) \, \frac{\Delta\xi_\Lambda^C(u,n \cdot p)}{\xi_\Lambda(n \cdot p)}\,, \nonumber \\
    &\Delta C_\gamma^A(q^2=0) = - \, g_{\rm em} \, M_{\Lambda_b}^2 \int \! du \left[C_3^C-C_4^C\right](n \cdot p,\bar{u}) \, \frac{\Delta\xi_\Lambda^C(u,q^2=0)}{\xi_\Lambda(q^2=0)}\,.
    \label{eq:newdeltaCA}
\end{align}

\section {Numerical results}
\label{section: numerical analysis}

Having established the SCET sum rules for the $\Lambda_b \to \Lambda$ form factor, we are now ready to investigate their phenomenological implications. We begin the numerical analysis by specifying the non-perturbative models for the $\Lambda_b$-baryon LCDA, determining the sum rule parameters, and evaluating the normalization constants $f_{\Lambda}(\nu)$ and $f_{\Lambda_b}^{(2)}(\mu)$. Finally, we present theoretical predictions for the $\Lambda_b \to \Lambda$ form factor at large hadronic recoil.

\subsection{Input parameters}

The LCDA of the $\Lambda_b$ baryon is derived in Appendix~\ref{App:DAs}. For our numerical calculations, we adopt the exponential model specified in \refeq{Lamb:LCDA}. In the context of the soft form factor, only the partially integrated function $\phi_4(\omega)$ contributes, and it takes a simple form at the soft scale:
$$
  \phi_4(\omega, \mu_0) := \frac{1}{\omega_0^2} \, e^{-\omega/\omega_0} \,.
$$

\begin{figure}
\begin{center}
\includegraphics[width=0.5 \columnwidth]{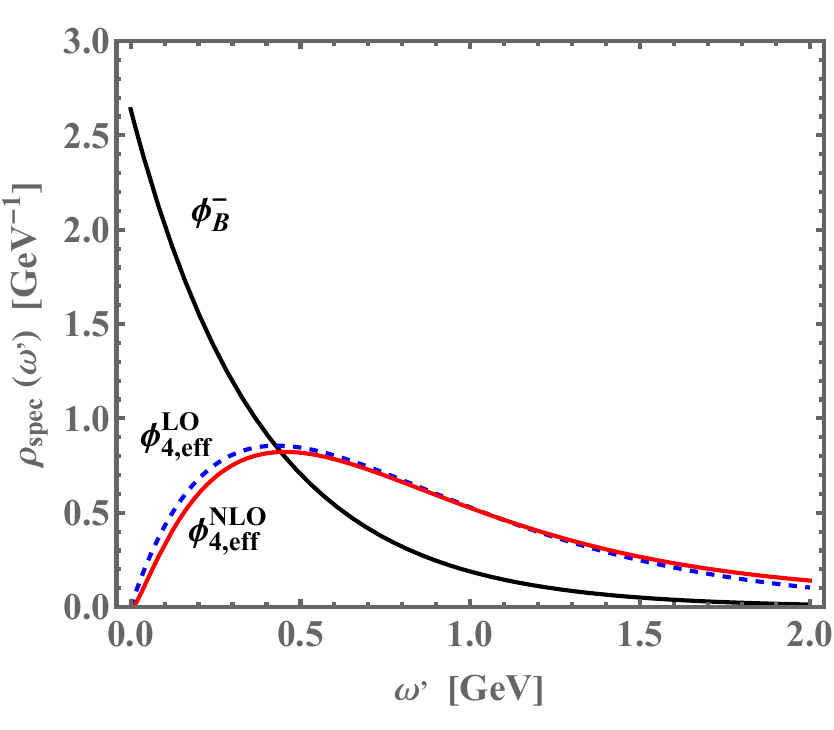}
\vspace*{0.01cm}
\caption{Spectral functions correspond to sum rules for the $B \rightarrow \pi$ form factors and $\xi_\Lambda$ for $\Lambda_b \to \Lambda$. The black solid curves represent the LO sum rule for the $B \rightarrow \pi$ form factors, which is simply $\phi_B^-(\omega^{\prime},1.5)$. The blue dashed and red solid curves show the LO and NLO sum rules for $\xi_\Lambda$, corresponding to $\phi_{4, \rm eff}^{\rm LO}(\omega^{\prime},1.5, 1.5)$ and $\phi_{4, \rm eff}^{\rm NLO}(\omega^{\prime},1.5, 1.5)$, respectively, with the central value $\omega_0=0.430$. }
\label{fig: SpecFunctions}
\end{center}
\end{figure}

Using the matching procedure from Ref.~\cite{Wang:2015ndk}, we take
$$
\xi_\Lambda^{\rm NLO}(q^2=0) \simeq f^+_{\Lambda_b \to \Lambda}(0) = 0.18 \pm 0.04
$$
as input to determine $\omega_0 = 430^{+70}_{-50} \,\, {\rm MeV}$. We then illustrate the effective ``distribution amplitude" $\phi_{4, \rm eff}(\omega^{\prime}, \mu, \nu)$ in \refeq{phi4eff} using this model in Fig.~\ref{fig: SpecFunctions}, 
with the central value $\omega_0 = 0.430~{\rm GeV}$. The blue dashed and red solid curves correspond to the sum rule predictions at LO and NLO, respectively.

\begin{figure}
\begin{center}
\includegraphics[width=1.0 \columnwidth]{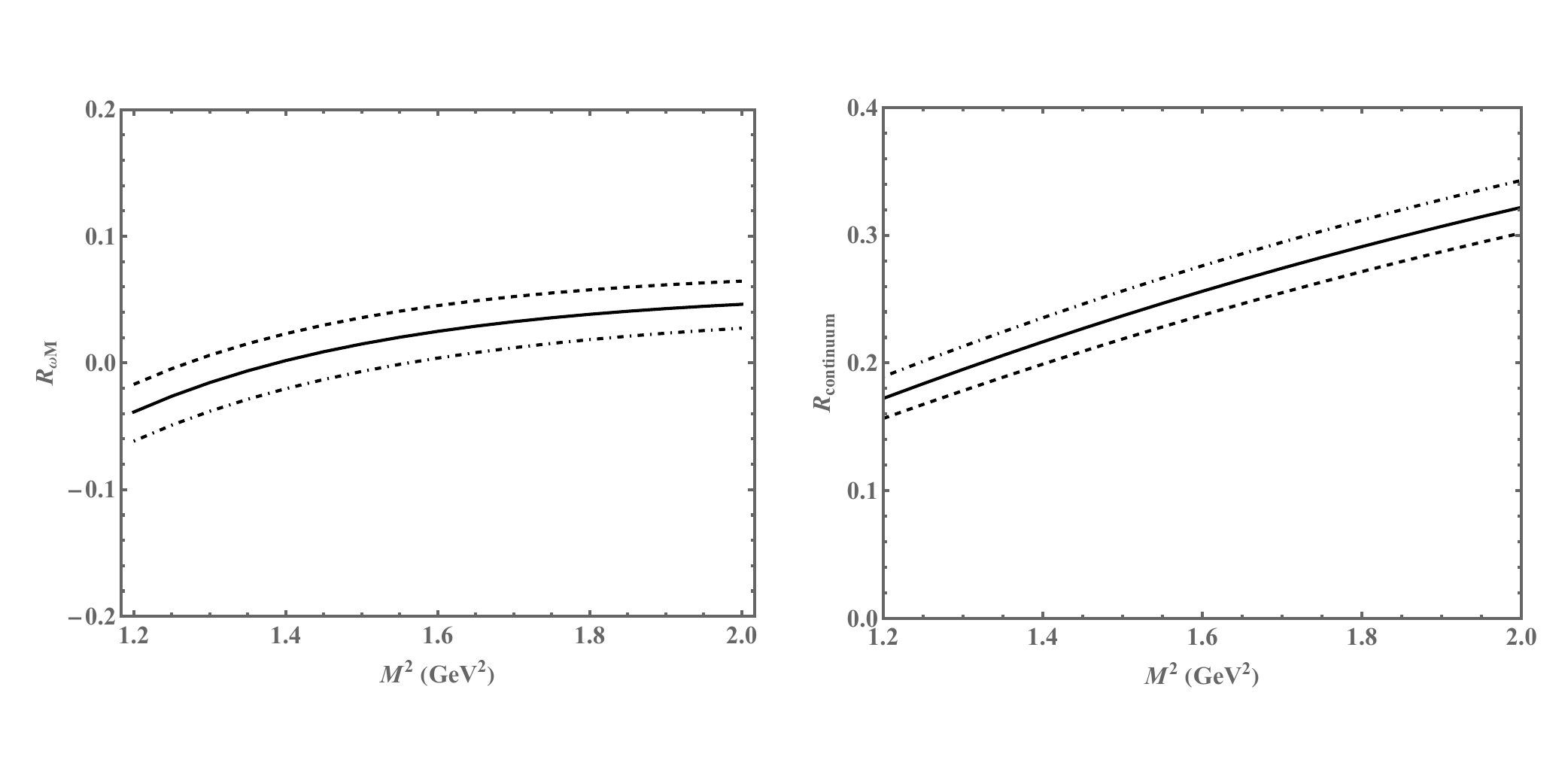}
\caption{Dependence of $R_{\omega_M}$ (left) and $R_{\rm continuum}$ (right) on the Borel mass.
Solid, dashed and dotted curves   correspond to the threshold $s_0=2.56 \, {\rm GeV^2}$,
$2.66 \, {\rm GeV^2}$,  $2.46 \, {\rm GeV^2}$, respectively, while all the other input parameters are fixed at their central values. }
\label{fig: Borelpara}
\end{center}
\end{figure}

Regarding the determination of the sum rule parameters, we follow the strategies
proposed for studying the sum rules of the $B \to \pi$ form factors~\cite{Wang:2015vgv}.
\begin{itemize}
\item To reduce the systematic uncertainty induced by the parton-hadron duality approximation,
the continuum contributions $R_{\rm continuum}$ to the correlation functions, displayed
in \refeq{ NLO correlator}, 
need to be reasonably controlled, i.e., kept below 40\%.
\item The sum rule predictions should be stable with respect to variations of the Borel mass parameter
$\omega_M$. More concretely, we impose the following condition on the logarithmic derivative of the form factor:
\begin{eqnarray}
R_{\omega_M} \equiv \frac{\partial \, \ln {\xi_ \Lambda}}{\partial \ln \omega_M}
\leq 40\% \,.
\end{eqnarray}
\end{itemize}
We show the dependencies of $R_{\omega_M}$ and $R_{\rm continuum}$ on the sum rule parameter $M^2$ in Fig.~\ref{fig: Borelpara}. 
The solid, dashed, and dotted curves correspond to the threshold values $s_0 = 2.56 \, {\rm GeV}^2$,
$2.66 \, {\rm GeV}^2$, and $2.46 \, {\rm GeV}^2$, respectively, while all other input parameters are fixed at their central values. 
The flatness of the curves demonstrates that our results exhibit a remarkably weak dependence on the Borel mass parameter.
Finally, the allowed ranges of the Borel parameter and the effective duality threshold are determined to be
\begin{eqnarray}
M^2 \equiv n \cdot p \, \omega_M = (1.6 \pm 0.4) \, {\rm GeV}^2 \,, \qquad
s_0 \equiv n \cdot p \, \omega_s = (2.56 \pm 0.10) \, {\rm GeV}^2 \,,
\end{eqnarray}
where the obtained interval for $s_0$ is in agreement with the values adopted in Refs.~\cite{Feldmann:2011xf,Liu:2014uha,Wang:2015ndk}.

We observe a significant discrepancy in the behavior of $R_{\omega_M}$ between our study and the $B \to \pi$ process. In the $B \to \pi$ scenario, the leading-order spectral function is given by $\phi^-_B(\omega^{\prime},\mu)$. As depicted in Fig.~\ref{fig: SpecFunctions}, its behavior at small $\omega^{\prime}$ differs markedly from that of $\phi_{4, \rm eff}^{\rm LO}(\omega^{\prime},\mu, \nu)$. Moreover, as illustrated in \refeq{NLOexact}, the mass of the final-state hadron appears in the exponential term. Given that the pion mass is substantially smaller than that of the $\Lambda$ baryon, this discrepancy in the small-$\omega^{\prime}$ regime contributes to the observed differences.
The dependence of $R_{\omega_M}$ on $M^2$ in baryon decay is negligible. In sharp contrast, $R_{\omega_M}$ exhibits a much stronger $M^2$ dependence in the $B \to \pi$ process.

In order to improve the convergence of the perturbative expansion, we need to resum the large logarithms to all orders in the perturbative matching coefficients by solving the RG evolution equations. However, following the argument of Ref.~\cite{Wang:2015vgv}, where it is noted that the characteristic scale of the jet function $\mu_{hc}$ is comparable to the hadronic scale $\mu_0$ entering the initial condition of the $\Lambda_b$-baryon DA, we will not resum the small logarithms of $\mu_{hc}/\mu_0$ arising from the RG running of the hadronic wave function when the factorization scale is chosen to be a hard-collinear scale of order $\sqrt{n \cdot p \, \Lambda_{\rm QCD}}$. Furthermore, the normalization parameter $f_{\Lambda_b}^{(2)}(\mu)$ will be taken directly from the HQET sum rule calculation; thus, no RG evolution of $f_{\Lambda_b}^{(2)}(\mu)$ is required \cite{Wang:2015ndk}.

The coupling $f_{\Lambda_b}^{(2)}(\mu_0)$ will be taken from the NLO HQET sum rule calculation, as given in Ref.~\cite{Groote:1997yr}:
\begin{eqnarray}
f_{\Lambda_b}^{(2)}(1 \, {\rm GeV}) = (3.0 \pm 0.5) \times 10^{-2} \, {\rm GeV^3} \,.
\end{eqnarray}
To reduce the theoretical uncertainty induced by the Borel mass parameter $\omega_M$, we employ the two-point QCD sum rules for the normalization parameter $f_{\Lambda}(\nu)$ \cite{Liu:2008yg}:
\begin{eqnarray}
f_{\Lambda}^2 \, e^{-m_{\Lambda}^2/M^2} &=& \frac{1}{640 \, \pi^4} \,
\int_{m_s^2}^{s_0} \, d s \,\, e^{-s/M^2} \, s \, \left(1 - \frac{m_s^2}{s} \right)^5 \nonumber \\
&& - \frac{1}{192 \, \pi^2} \, \left\langle \frac{\alpha_s}{\pi} GG \right\rangle \,
\int_{m_s^2}^{s_0} \, d s \, e^{-s/M^2} \, \frac{m_s^2}{s^2} \, \left(1 - \frac{m_s^2}{s} \right) \,
\left(1 - \frac{2 \, m_s^2}{s} \right) \,,
\end{eqnarray}
evaluated at tree level. For the numerical analysis, we use the gluon condensate density
$\left\langle \alpha_s / \pi \, GG \right\rangle = \left(1.2^{+0.6}_{-1.2} \right) \times 10^{-2} \,\, {\rm GeV^4}$.

For the resummation of large logarithms in the hard functions, we solve the RG equation for the coefficients of the A-type currents, given in \refeq{RGE of hard functions}. The solution reads \cite{Hill:2004if}
\begin{align}
C_i^A(n\cdot p, \mu) = \left(\frac{n \cdot p}{\mu_h} \right)^{a(\mu_h, \mu)} 
\exp \left[ S(\mu_h, \mu) + \frac{5 C_F}{2 \beta_0} \ln r_1 \right] \, C_i^A(n\cdot p, \mu_h) \,,
\end{align}
with \cite{Bosch:2003fc}
\begin{align}
S(\mu_h, \mu) &= \frac{\Gamma_0}{4 \beta_0^2} \left[ \frac{4\pi}{\alpha_s(\mu_h)} \left(1 - \frac{1}{r_1} - \ln r_1 \right) + \frac{\beta_1}{2\beta_0} \ln^2 r_1 \right. \nonumber \\
&\left. \qquad - \left( \frac{\Gamma_1}{\Gamma_0} - \frac{\beta_1}{\beta_0} \right) (r_1 - 1 - \ln r_1) \right] \,, \nonumber \\
a(\mu_h, \mu) &= -\frac{\Gamma_0}{2 \beta_0} \ln r_1 \,,
\end{align}
where $r_1 = \alpha_s(\mu) / \alpha_s(\mu_h) \geq 1$. For the relevant coefficients entering the cusp anomalous dimension and the $\beta$ function, we take $n_f = 4$.

We now turn to the discussion of the choices for the renormalization and factorization scales entering the LL sum rules. The renormalization scale $\nu$ of the baryonic current and the factorization scale $\mu$ will be varied within the interval
$1\,{\rm GeV} \leq \mu, \nu \leq 2\,{\rm GeV}$ around the default value $\mu = \nu = 1.5\,{\rm GeV}$.
The hard scale $\mu_h$ appearing in the hard matching coefficients will be set to $\mu_h = m_b$, with variations in the range $[m_b/2,\ 2 m_b]$.
For the $\Lambda_b \to \Lambda \gamma$ process, due to the specific logarithmic structure of the hard coefficient in \refeq{CgammaA}, the hard scale will instead be taken as $\mu_h = m_b^{\rm pole}$, varied within the range $[m_b,\ M_{\Lambda_b}]$.
In addition, we adopt the $\overline{\rm MS}$ bottom-quark mass 
$\overline{m_b}(\overline{m_b}) = 4.183 \pm 0.007\,{\rm GeV}$ and the strange-quark mass $\overline{m_s}(\overline{m_s}) = 0.0935 \pm 0.0008\,{\rm GeV}$ \cite{ParticleDataGroup:2024cfk}.
The pole masses of the $b$- and $c$-quarks are deduced from the PS mass \cite{Beneke:1998rk,Shen:2020hfq}.

\subsection{Predictions for the \texorpdfstring{$\Lambda_b \to \Lambda$}{Lambda\_b → Lambda} effective form factors}
\label{sec:formfactornum}

\begin{figure}[t]
\centering
\includegraphics[width=.8 \columnwidth]{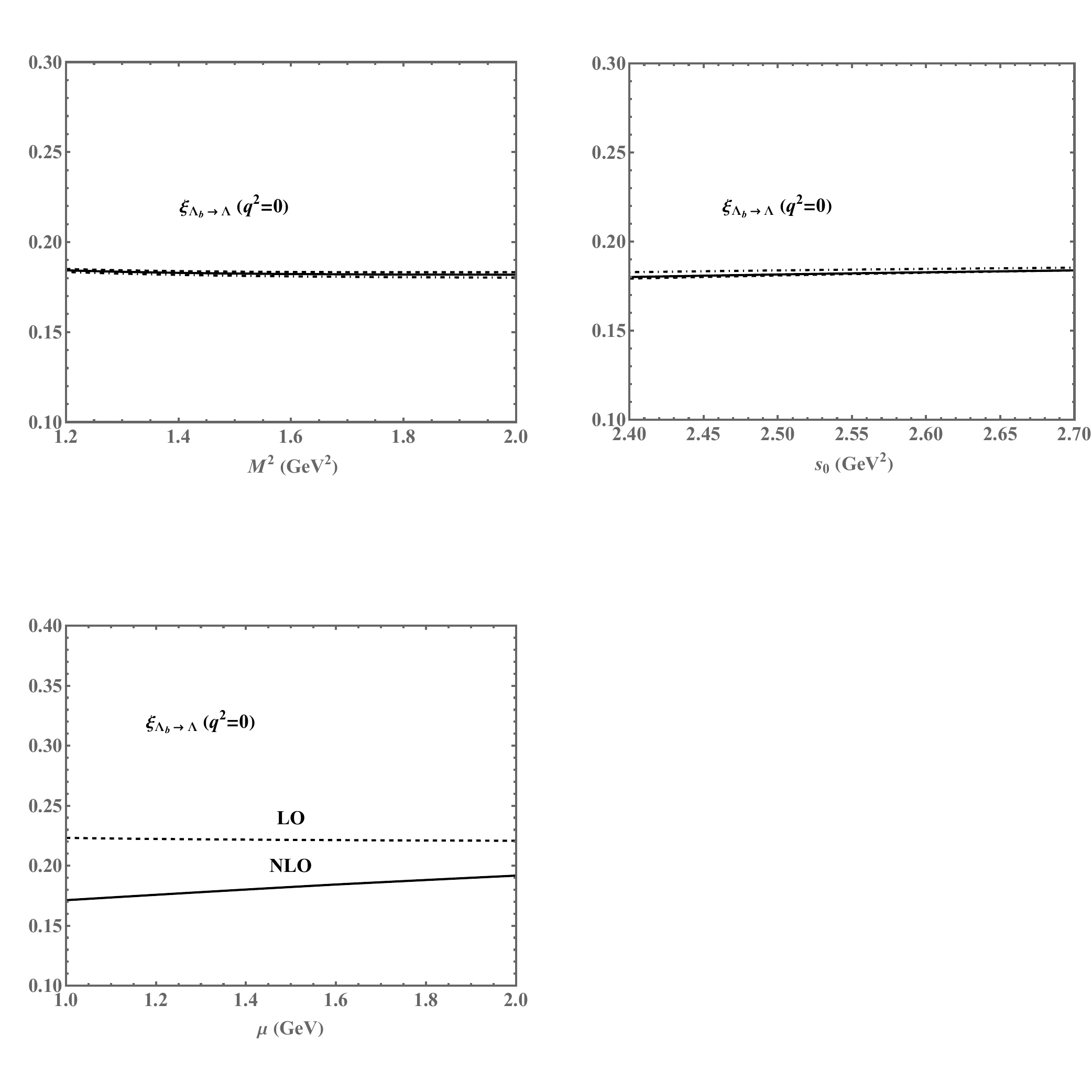}
\vspace*{0.1cm}
\caption{Dependence of $\xi_{\Lambda}(q^2=0)$ on the Borel parameter (top left),
on the threshold parameter (top right) and on the factorization scale (bottom left).
Solid, dashed and dotted-dashed curves are obtained from the sum rules with $s_0=2.56 \, {\rm GeV^2}$,
$2.66 \, {\rm GeV^2}$,  $2.46 \, {\rm GeV^2}$ (top left) and $M^2=1.6  \, {\rm GeV^2}$,
$2.0 \, {\rm GeV^2}$, $1.2 \, {\rm GeV^2}$ (top right), respectively, while all the other input parameters
are fixed at their central values. The curves labelled by  ``LO'' and ``NLO" (bottom)
correspond to the sum rule predictions at LO and NLO accuracy. }
\label{fig: sum rule parameter and mu dependence of FF}
\end{figure}

To illustrate key numerical features of the LCSR predictions, we present the dependencies of 
$\xi_{\Lambda}(q^2=0)$ on the sum rule parameters $M^2$ and $s_0$, as well as on the factorization 
scale $\mu$, in Fig.~\ref{fig: sum rule parameter and mu dependence of FF}. It is evident that the 
sum rule for $\xi_{\Lambda}(q^2=0)$ exhibits exceptionally mild dependence on the Borel mass parameter, 
owing to a strong cancellation of systematic uncertainties between the LCSR prediction for 
$\xi_{\Lambda}(q^2=0)$ and the QCD sum rule for the coupling $f_{\Lambda}$. 
Also the three different choices of $s_0$ introduce only a negligible difference to the form factor.
 In our case, the NLO sum rules 
show sensitivity to the factorization scale $\mu$, since SCET form factors do not involve hard matching 
coefficients as in the QCD LCSR framework.

\begin{figure}[t]
\begin{center}
\includegraphics[width=.9 \columnwidth]{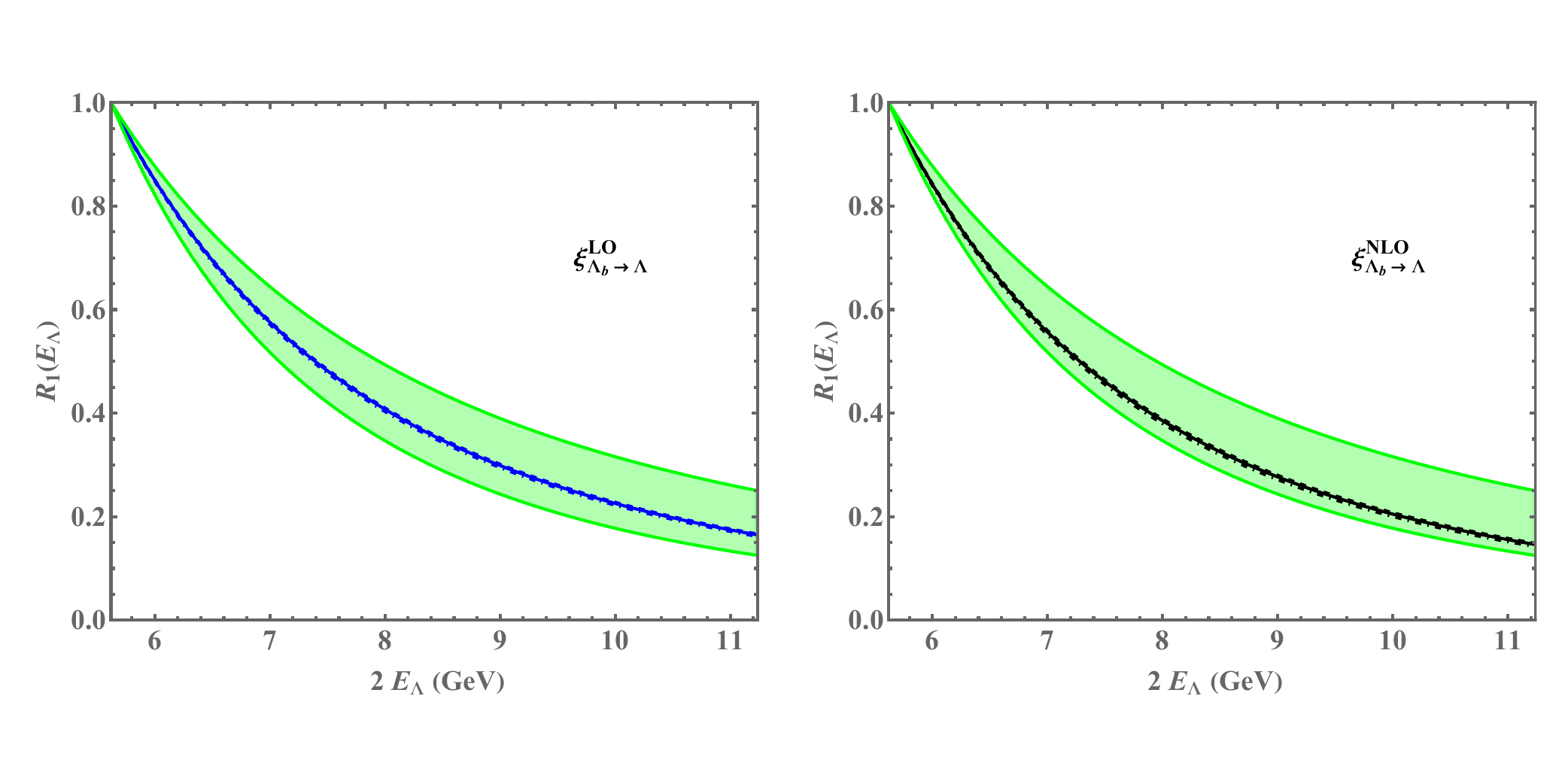}
\caption{Dependence of the ratio $R_1(E_{\Lambda})$ on the $\Lambda$-baryon energy $E_{\Lambda}$.
The blue (left panel) and the black (right panel) curves  are obtained from the LO and NLO sum rule predictions,
respectively. The two green curves refer to a pure $1/E_{\Lambda}^2$ and a pure $1/E_{\Lambda}^3$
dependence.}
\label{fig: energy dependence of FF}
\end{center}
\end{figure}

We now turn to investigate the $\Lambda$-baryon energy dependence of the form factor
$\xi_{\Lambda}(n \cdot p)$ using the sum rules at LO and NLO accuracy. This study is of particular conceptual interest since the soft-overlap contributions
and the hard-spectator scattering effects in heavy-to-light baryonic form factors
exhibit different scalings with respect to $1/E_{\Lambda}$ at large hadronic recoil.
To this end, we introduce the ratio originally proposed in Ref.~\cite{DeFazio:2007hw}:
\begin{eqnarray}
R_1(E_{\Lambda}) = \frac{\xi_{\Lambda}(n \cdot p)}{\xi_{\Lambda}(m_{\Lambda_b})} \,.
\end{eqnarray}
As shown in Fig.~\ref{fig: energy dependence of FF},
the predicted energy dependence of $\xi_{\Lambda}$
from the LO and NLO sum rules displays a scaling behavior between $1/E_{\Lambda}^2$ and $1/E_{\Lambda}^3$ for
the default choices of theory input parameters.

As discussed in Sec.~\ref{sec: LP}, the leading-power contribution to $\xi_\Lambda$ is ${\cal O}(\alpha_s^2)$, with the numerical estimate \cite{Wang:2011uv}
\begin{equation}
    \xi_{\Lambda}^{\rm LP}(q^2=0) = -0.012^{+0.009}_{-0.023} \,.
    \label{eq:numLP}
\end{equation}
The soft contributions calculated in this work are
\begin{equation}
    \xi_{\Lambda}^{\rm LO}(q^2=0) = 0.222^{+0.057}_{-0.058} \,, \quad
    \xi_{\Lambda}^{\rm NLO}(q^2=0) = 0.182^{+0.050}_{-0.041} \,.
    \label{eq:num-soft}
\end{equation}
The dominant theoretical uncertainty for the form factor at $q^2=0$ computed from the SCET LCSR arises from the variation of the parameter $\omega_0$ entering the $\Lambda_b$-baryon DA $\phi_4(\omega, \mu_0)$. 
Comparing Eqs.~\eqref{eq:numLP} and \eqref{eq:num-soft}, we observe that the A-type form factor is numerically dominated by soft contributions, despite being formally power-suppressed. This observation is consistent with the conclusions of Ref.~\cite{Wang:2011uv}.

To estimate the non-factorizable contributions, we also calculate the C-type form factor $\Delta\xi^C_\Lambda(u, n\cdot p)$. To preserve quark-hadron duality, we restrict our analysis to $q^2$ values exceeding the masses of narrow vector resonances like the $\rho$ and $\omega$ mesons \cite{Feldmann:2023plv}. 
Accordingly, we evaluate our LCSRs at the following $q^2$ points: $q^2 = \{2,4,6\} \,\mathrm{GeV}^2$,  
\begin{equation}
\begin{aligned}
    10^{4} \cdot \int du\, \Delta\xi^C_\Lambda(u, q^2=2\,\mathrm{GeV}^2) &= -(1.06 \pm 0.19) - i \,(2.49 \pm 0.77) \,, \\
    10^{4} \cdot \int du\, \Delta\xi^C_\Lambda(u, q^2=4\,\mathrm{GeV}^2) &= (0.45 \pm 0.40) - i \,(2.29 \pm 0.54) \,, \\
    10^{4} \cdot \int du\, \Delta\xi^C_\Lambda(u, q^2=6\,\mathrm{GeV}^2) &= (1.09 \pm 0.42) - i \,(1.58 \pm 0.28) \,.
\end{aligned}
\end{equation}
The non-factorizable contributions to the $\Lambda_b \to \Lambda \ell^+ \ell^-$ and $\Lambda_b \to \Lambda \gamma$ decay amplitudes can be expressed as shifts in the hard functions $C_1^A$ and $C_\gamma^A$, as given in \refeq{newdeltaCA}. For $\Lambda_b \to \Lambda \gamma$ decay, we find 
\begin{equation}
  \frac{|\Delta C_\gamma^A|}{|C_\gamma^A|} < 2\% 
\end{equation}
at the scale $\mu = m_b^{\rm pole}$. In the case of $\Lambda_b \to \Lambda \ell^+ \ell^-$ decay, the ratio $|\Delta C_1^A|/|C_1^A|$ is typically $\mathcal{O}(1\%)$ across most of the region $1 \, \mathrm{GeV}^2 < q^2 < 7 \, \mathrm{GeV}^2$. However, a notable peak appears around $q^2 \sim 3.2 \, \mathrm{GeV}^2$, where
\begin{equation}
  \frac{|\Delta C_1^A(q^2 \sim 3.2 \, \mathrm{GeV}^2)|}{|C_1^A(q^2 \sim 3.2 \, \mathrm{GeV}^2)|} \sim 50\% \,.
\end{equation}
This peak arises because both the operators $\mathcal{Q}_7$ and $\mathcal{Q}9$ contribute to $C_1^A$, and their effects largely cancel each other near $q^2=3.2 \, \mathrm{GeV}^2$. As a result, at the amplitude level, the operator $\mathcal{Q}_{10}$ dominates in this region. In full QCD, operators $\mathcal{Q}_7$ and $\mathcal{Q}_9$ will contribute to different form factors, and the non-factorizable effects can be interpreted as a shift in $C_9$. As demonstrated in Ref.~\cite{Feldmann:2023plv}, the relative shift $|\Delta C_{9,\perp}|/|C_9|$ is only $\mathcal{O}(1\%)$. These results indicate that the non-factorizable contributions to the branching ratios remain relatively small.
Moreover, the one-loop matching of C-type operators from QCD to $\scetone$ is incomplete, and the renormalization group equations for these operators are not yet available. Therefore, we neglect the non-factorizable contributions in our phenomenological analysis and defer their detailed study to future work.

\section{Phenomenological applications}
\label{section: phenomenologies}

In this section, we explore the phenomenological applications of the calculated $\xi_\Lambda$ form factor, which serves as a fundamental ingredient in the theoretical description of the electroweak penguin–induced $\Lambda_b \to \Lambda \, \ell^{+} \ell^{-}$ and $\Lambda_b\to\Lambda \, \gamma$ decays. As discussed in Sec.~\ref{sec:formfactornum}, we restrict ourselves to the factorizable contributions to the decay amplitudes.

\begin{figure}[htbp]
\begin{center}
\includegraphics[width=1.0 \columnwidth]{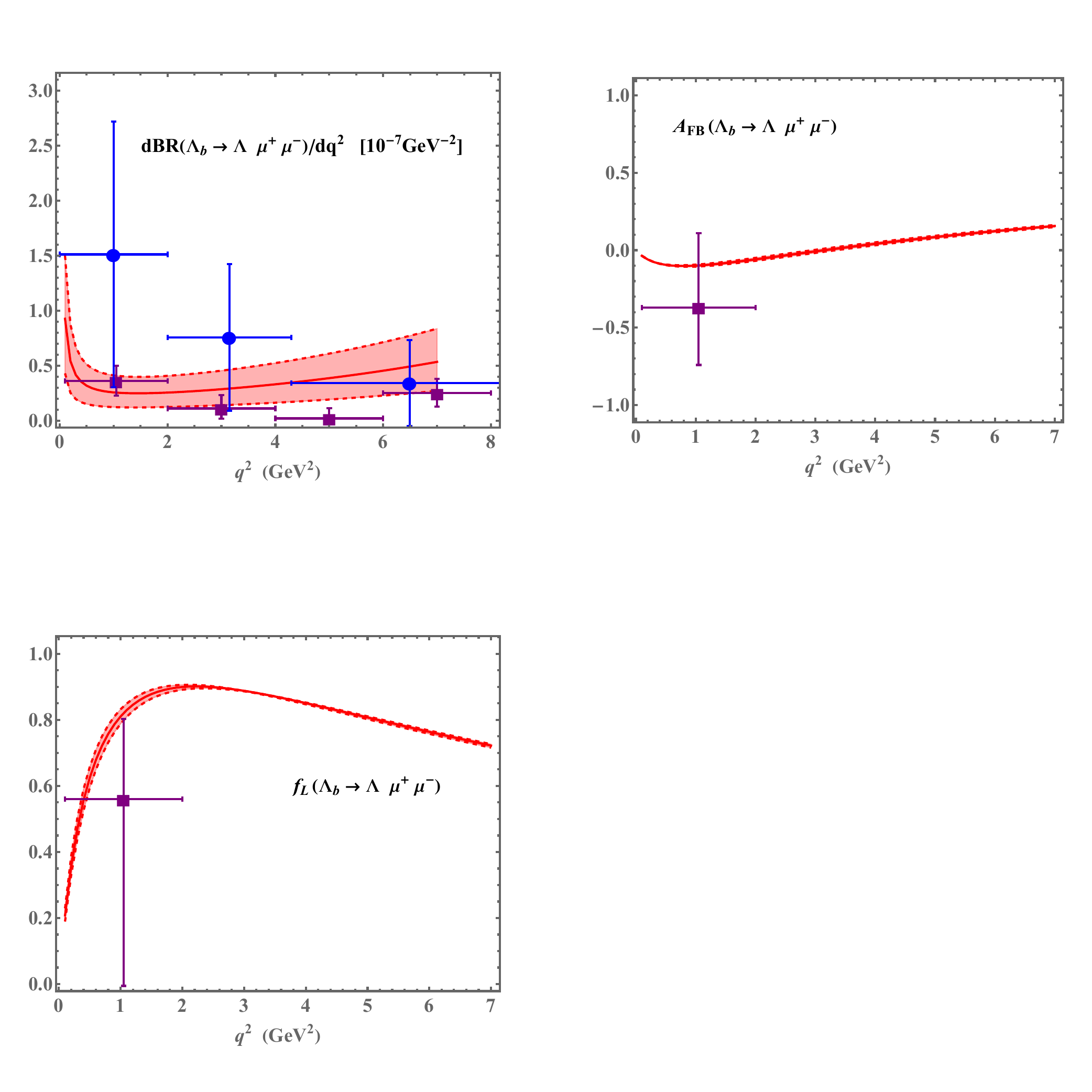}
\vspace*{0.1cm}
\caption{The differential branching fraction,
the leptonic forward-backward asymmetry and the longitudinal polarization fraction
of the di-lepton system for $\Lambda_b \to \Lambda \, \ell^{+} \ell^{-}$  as functions of $q^2$
in the factorization limit. The solid (pink) curve corresponds to the LL sum rule predictions
with  the shaded region (pink) indicating the theory uncertainties from form factor. The experimental data bins are taken from LHCb \cite{LHCb:2015tgy}
(purple squares) and CDF  \cite{CDF:2012qwd} (blue full circles). }
\label{fig: differential observables}
\end{center}
\end{figure}

The double differential decay distribution of $\Lambda_b \to \Lambda \, \ell^{+} \ell^{-}$, in terms of the momentum transfer squared $q^2$ and the angle $\theta$ between the negatively charged lepton and the $\Lambda$ baryon in the rest frame of the lepton pair, is derived in Appendix \ref{App: helicity amplitude}:
\begin{eqnarray}
\frac{d \Gamma(\Lambda_b \to \Lambda \, \ell^{+} \ell^{-})}{d q^2 \, d \cos \theta}
= {\cal N} \, \left[ F_T(q^2) \left( 1 + \cos^2 \theta \right)
+ 2 \, F_A(q^2) \, \cos \theta + 2 \, F_L(q^2) \left( 1 - \cos^2 \theta \right) \right] .
\end{eqnarray}
Evaluating the helicity amplitudes with the form factor computed from the LL LCSR obtained above yields the differential branching fraction of $\Lambda_b \to \Lambda \, \ell^{+} \ell^{-}$ as a function of $q^2$, shown in Fig.~\ref{fig: differential observables}. The theory predictions are also compared with experimental measurements from CDF \cite{CDF:2012qwd} and LHCb \cite{LHCb:2015tgy}. The LHCb data are systematically lower than the theory predictions at large recoil, while the sizable uncertainties in the CDF measurements, combined with the large theoretical uncertainties at high $q^2$, prevent drawing a definite conclusion.


Following Ref.~\cite{LHCb:2015tgy}, we further consider the forward-backward asymmetry and the longitudinal polarization fraction of the di-lepton system:
\begin{eqnarray}
A_{\rm FB}(q^2) &=& \frac{ \int_{-1}^{0} \! d \cos \theta \,
\frac{d \Gamma(\Lambda_b \to \Lambda \, \ell^{+} \ell^{-})}{d q^2 \, d \cos \theta}
- \int_{0}^{1} \! d \cos \theta \,
\frac{d \Gamma(\Lambda_b \to \Lambda \, \ell^{+} \ell^{-})}{d q^2 \, d \cos \theta} }
{\frac{d \Gamma(\Lambda_b \to \Lambda \, \ell^{+} \ell^{-})}{d q^2}} \,, \nonumber \\
f_L(q^2) &=& \frac{F_L(q^2)}{F_L(q^2) + F_T(q^2)} \,, \label{eq:asymmetry}
\end{eqnarray}
where the definition of $A_{\rm FB}(q^2)$ differs from Refs.~\cite{Feldmann:2011xf,Wang:2015ndk} due to a distinct convention for the angle $\theta$.
We also plot the $q^2$ dependence of the differential forward-backward asymmetry and the longitudinal polarization fraction in Fig.~\ref{fig: differential observables}. It is apparent that the theoretical uncertainties for these two observables are much smaller than those for the branching ratios, owing to partial cancellation of uncertainties in the ratios defined in \refeq{asymmetry}. However, as seen in Fig.~\ref{fig: differential observables}, the experimental uncertainties for these observables remain quite large. We expect future data to shed more light on the rich phenomenology encoded in the angular distributions of $\Lambda_b \to \Lambda \, \ell^{+} \ell^{-}$, with an emphasis on optimized physics observables that are especially sensitive for new physics searches.

With the soft form factor $\xi_\Lambda$ at $q^2 = 0$ in hand, we further analyze the $\Lambda_b \to \Lambda \gamma$ decay in Appendix \ref{App: helicity amplitude}. Employing \refeq{branchingratio}, we obtain
\begin{equation}
\mathcal{B}(\Lambda_b \to \Lambda \gamma) = (1.22^{+0.66}_{-0.67}) \times 10^{-5},
\end{equation}
while the experimental value, reported in \cite{ParticleDataGroup:2024cfk}, is $(7.1 \pm 1.7) \times 10^{-6}$. Our analysis shows that the central theoretical prediction is slightly larger than the experimental measurement. However, it is important to emphasize that the experimental value lies within the theoretical uncertainty band, and both theoretical and experimental uncertainties remain sizable. Considering both experimental and theoretical uncertainties, our prediction is consistent with the current data. In our future work, reducing theoretical errors will be one of our primary tasks. On one hand, within the framework of LCSRs, we can minimize the input parameters such as that in the $\Lambda_b$ distribution amplitudes. On the other hand, we can conduct joint fits by combining lattice simulation results of transition form factors and our calculations.

\section{Conclusions}
\label{section: conclusion}

In this work, we study the $\Lambda_b \to \Lambda \, \ell^{+} \ell^{-}$ and $\Lambda_b \to \Lambda \, \gamma$ decays using light-cone sum rules within the soft-collinear effective theory framework. To obtain the decay amplitudes, we parameterize the matrix elements of $\scetone$ operators in terms of effective form factors. Employing light-cone sum rules, we perform the perturbative matching from $\scetone$ to $\scettwo$ to determine the hard-collinear functions in the SCET factorization framework.

For the A-type effective form factor $\xi_\Lambda$, we obtain the soft form factor up to NLO accuracy, and the resulting jet function matches that derived using LCSR combined with the method of regions.
The leading-power contribution of the form factor is $\alpha^2_s$ suppressed compared to the soft form factor.
The form factor $\Delta\xi^B_\Lambda$ represents a non-local form factor associated with hard-scattering corrections mediated by hard-collinear gluon exchange. This form factor vanishes at leading power in the relevant correlation functions, indicating that the hard-scattering corrections are power-suppressed relative to the ``soft'' contribution. This behavior is distinct from that observed in $B$-meson decays.
The C-type form factors $\Delta_k\xi^C_\Lambda$ also represent non-local form factors corresponding to non-factorizable contributions. To calculate the hard coefficients of the C-type $\scetone$ operators, we consider not only the weak annihilation contribution but also quark-loop and chromomagnetic contributions. We then carry out the $\scetone$ $\to$ $\scettwo$ matching of the $\gamma^\ast$-to-$\Lambda_b$ correlation functions to derive these form factors, and find $\Delta_1\xi^C_\Lambda = \Delta_2\xi^C_\Lambda = 0$ and $\Delta_3\xi^C_\Lambda = -\Delta_4\xi^C_\Lambda$.
At the amplitude level, we find that the C-type non-factorizable contributions are numerically suppressed, indicating that the $\Lambda_b$ decay processes are dominated by the ``soft" contributions. 

We compute the $q^2$-dependent differential branching fractions in the factorization limit, the forward-backward asymmetries, and the dilepton longitudinal polarization fractions for $\Lambda_b \to \Lambda \ell^+\ell^-$ decays, as well as the branching fractions for $\Lambda_b \to \Lambda \gamma$ decays. The theoretical predictions for the branching fractions carry substantial uncertainties, dominated by the poorly constrained $\Lambda_b$ LCDAs. This situation underscores the urgent need for improved determinations of these non-perturbative parameters in future investigations (see recent developments in Refs. \cite{LatticeParton:2024vck,Feldmann:2025dcs}). In striking contrast, the forward-backward asymmetry and dilepton longitudinal polarization fraction exhibit remarkably small theoretical uncertainties, benefiting from significant cancellation of parametric errors in their ratio definitions.

In summary, this theoretical investigation establishes a robust framework for analyzing strong interaction effects in exclusive $\Lambda_b$-baryon decays within the SCET formalism. Our results not only provide crucial insights into the factorization properties and power counting of these processes, but also lay the groundwork for future precision studies. Given the remarkable progress in experimental capabilities for beauty baryon decays at LHCb and future colliders, we anticipate that this work will inspire several important extensions, including calculations of higher-order corrections, explorations of additional decay channels, and more comprehensive analyses incorporating improved determinations of baryon distribution amplitudes.

\section*{Acknowledgements}
We thank Jian-Peng Wang for his help to the derivation of angular distributions.
This work is supported in part by the National Key Research and Development Program of China (2023YFA1606000), Natural Science Foundation of China under
grant No.12175218, 12275277, 12305099 and 12435004 and the Natural Science Foundation of Shandong Province under the Grant No. ZR2024MA076.

\appendix

\section{Light-Cone Distribution Amplitudes}

\label{App:DAs}

The most general non-local matrix elements for $\Lambda_b$ baryon in coordinate space is~\cite{Feldmann:2011xf}:
\begin{align}
 & \epsilon_{ijk} \, \langle 0| \left(u^i_\alpha (z_1) \,  d^j_\beta(z_2)\right) h^k_{v}(0)
  |\Lambda_b(v,s)\rangle
  \cr
 \equiv & \frac14 \left \{ f^{(1)}_{\Lambda_b} \left[ \tilde M^{(1)}(v,z_1,z_2) \gamma_5 C^T\right]_{\beta\alpha}
 +
f^{(2)}_{\Lambda_b} \left[ \tilde M^{(2)}(v,z_1,z_2) \gamma_5 C^T\right]_{\beta\alpha} \right\} u_{\Lambda_b}(v,s) \,,
\end{align}
with a part that contains an odd number of Dirac matrices $\tilde M^{(2)}(v,z_1,z_2)$, and a part that contains an even number of Dirac matrices $\tilde M^{(1)}(v,z_1,z_2)$.

The standard LCDAs can be obtained by expanding the above expression around the limit $z_1^2=z_2^2=z_1\cdot z_2=0$ (correspond to multiple expand in SCET). As shown in Ref.\cite{Bell:2013tfa}, in this limit, $\bar n \cdot z_i\ll z_i^\perp \ll n \cdot z_i$,
\begin{align}
\tilde M^{(2)}(v,z_1,z_2) & \longrightarrow
\frac{\slashed n}{2} \, \tilde \phi_2(\tau_1,\tau_2)
+
\frac{\slashed {\bar n}}{2} \left( \tilde \phi_2(\tau_1,\tau_2)+\tilde \phi_{42}^{(i)}(\tau_1,\tau_2) + \tilde \phi_{42}^{(ii)}(\tau_1,\tau_2) \right)
\cr & \quad
+
\frac{\tilde \phi_{42}^{(i)}(\tau_1,\tau_2)}{2\tau_1} \,  \slashed z_1^\perp
+
\frac{\tilde \phi_{42}^{(ii)}(\tau_1,\tau_2)}{2\tau_2} \,  \slashed z_2^\perp
\cr
& \quad
+
\tilde\phi_X(\tau_1,\tau_2) \left( \frac{\slashed z_1^\perp}{2\tau_1} - \frac{\slashed z_2^\perp}{2\tau_2}
 \right) \left(\frac{\slashed {\bar n} \slashed n}{4} - \frac{\slashed n \slashed {\bar n}}{4} \right)
 + {\cal O}(z_{i\perp}^2, \bar n \cdot z_i) \,,
 \label{eq:M2zexp}
\end{align}
where we denote with $\tau_i = \frac{n \cdot z_i}{2}$
the Fourier-conjugate variables to the momentum components
$\omega_i = \bar n \cdot k_i$ of the associated light-quark states in the heavy baryon. After Fourier transformation, the general momentum-space representation for \refeq{M2zexp}
including the first-order terms off the light-cone (in $D$ dimensions) becomes
\begin{align}
M^{(2)}(\omega_1,\omega_2) = & \  \frac{\slashed n}{2} \, \phi_2(\omega_1,\omega_2)
+
\frac{\slashed {\bar n}}{2} \, \phi_4(\omega_1,\omega_2)
\cr
&  - \frac{1}{D-2} \, \gamma_\mu^\perp \, \int_0^{\omega_1} d\eta_1 \left( \phi_{42}^{(i)}(\eta_1,\omega_2) - \phi_X(\eta_1,\omega_2) \right)
\frac{\slashed n\slashed {\bar n}}{4}
\, \frac{\partial}{\partial k_{1\mu}^\perp}
\cr & - \frac{1}{D-2} \, \gamma_\mu^\perp \, \int_0^{\omega_1} d\eta_1 \left( \phi_{42}^{(i)}(\eta_1,\omega_2) + \phi_X(\eta_1,\omega_2) \right)
\frac{\slashed {\bar n}\slashed n}{4}
\, \frac{\partial}{\partial k_{1\mu}^\perp}
\cr & - \frac{1}{D-2} \, \gamma_\mu^\perp \, \int_0^{\omega_2} d\eta_2 \left( \phi_{42}^{(ii)}(\omega_1,\eta_2) - \phi_X(\omega_1,\eta_2) \right)
\frac{\slashed {\bar n}\slashed n}{4}
\, \frac{\partial}{\partial k_{2\mu}^\perp}
\cr & - \frac{1}{D-2} \, \gamma_\mu^\perp \, \int_0^{\omega_2} d\eta_2 \left( \phi_{42}^{(ii)}(\omega_1,\eta_2) + \phi_X(\omega_1,\eta_2) \right)
\frac{\slashed n\slashed {\bar n}}{4}
\, \frac{\partial}{\partial k_{2\mu}^\perp} \,.
 \label{M2proj}
\end{align}
In the Wandzura-Wilczek approximation,
\begin{align}
 \int_0^{\omega_1} d\eta_1 \left( \phi_{42}^{(i)}(\eta_1,\omega_2) + \phi_X(\eta_1,\omega_2) \right)
= \omega_1 \, \phi_4(\omega_1,\omega_2) \,,
\cr
 \int_0^{\omega_2} d\eta_2 \left( \phi_{42}^{(ii)}(\omega_1,\eta_2) + \phi_X(\omega_1,\eta_2) \right)
= \omega_2 \, \phi_4(\omega_1,\omega_2) \,.
\label{eq: WW}
\end{align}
With the most general form of the momentum-space projectors, we can write 
\begin{align}
 \phi_4(\omega_1,\omega_2)&=
 \int_{\omega_1}^\infty dx_1 \int_{\omega_2}^\infty dx_2 \,
  (x_1-\omega_1)(x_2-\omega_2) \, \psi_v(x_1,x_2) \,.
\end{align}
In the simplest case, we could   model the wave functions by assuming an exponential dependence of $\psi_v$
on $(x_1+x_2)$ with a single hadronic parameter $\omega_0$ measuring the average energy of the light quarks,
\begin{align}
 \psi_v(x_1,x_2) & \to \frac{\exp \left( - \frac{x_1+x_2}{\omega_0} \right)}{\omega_0^6} \,.
\end{align}
This ansatz yields the following exponential model,
\begin{align}
 \phi_4(\omega_1,\omega_2) \to \frac{1}{\omega_0^2} \, e^{-(\omega_1+\omega_2)/\omega_0} \,.
\label{eq:Lamb:LCDA}
\end{align}

However, when we calculate the non-factorizable contributions in Fig.~\ref{fig: NF}, we rather have to consider the expansion for the situation 
$n\cdot z_1 \ll z_1^\perp \ll \bar n \cdot z_1$ and $\bar n \cdot z_2 \ll z_2^\perp \ll n\cdot z_2$, 
such that $t_1 \approx \bar \tau_1 = \frac{\bar n \cdot z_1}{2}$, $t_2 \approx \tau_2 = \frac{n \cdot z_2}{2}$, and 
$z_1 \cdot z_2 \approx 2 \bar \tau_1 \tau_2$. For completeness, we adopt the methodology established in Refs. \cite{Bell:2013tfa,Feldmann:2023plv} to derive the momentum projectors in this case. The momentum-space projector for matrix $\tilde M^{(1)}$ takes the form:
\begin{align}
    M^{(1)}(\bar{\omega}_1,\omega_2) 
    & =   
    \frac{\slashed n\slashed {\bar n}}{4} \, 
    \left(
        \chi_{3}^{(0)}(\bar{\omega}_1,\omega_2)
        -
        \chi_{Y}(\bar{\omega}_1,\omega_2)
    \right)
    \cr
    & +
    \frac{\slashed {\bar n} \slashed n}{4} 
    \left( 
        \chi_{3}^{(0)}(\bar{\omega}_1,\omega_2)
+
        \chi_{Y}(\bar{\omega}_1,\omega_2)
 +
        \chi_{3}^{(i)}(\bar{\omega}_1,\omega_2)
        +
        \chi_{3}^{(ii)}(\bar{\omega}_1,\omega_2) 
    \right)
    \cr
    & + 
    \frac{1}{2} \, \gamma_\rho^\perp \,
        \bar\chi_{3}^{(i)}(\bar{\omega}_1,\omega_2) \,
    \frac{\slashed n}{2}
    \, \frac{\partial}{\partial k_{1\rho}^\perp}
    \cr 
    & +
    \frac{1}{2} \, \gamma_\rho^\perp
    \left( 
        \bar\chi_{3}^{(i)}(\bar{\omega}_1,\omega_2) + 2\bar\chi_Y(\bar{\omega}_1,\omega_2) 
    \right)
    \frac{\slashed {\bar n}}{2}
    \, \frac{\partial}{\partial k_{1\rho}^\perp}
    \cr 
    & - 
    \frac{1}{2} \, \gamma_\rho^\perp \,
\hat\chi_{3}^{(ii)}(\bar{\omega}_1,\omega_2) \,
    \frac{\slashed {\bar n}}{2}
    \, \frac{\partial}{\partial k_{2\rho}^\perp}
    \cr
    & -
    \frac{1}{2} \, \gamma_\rho^\perp
    \left( 
        \hat\chi_{3}^{(ii)}(\bar{\omega}_1,\omega_2) + 2\hat\chi_Y(\bar{\omega}_1,\omega_2) 
    \right)
    \frac{\slashed n}{2}
    \, \frac{\partial}{\partial k_{2\rho}^\perp} \,.
    \label{eq:newlcproj1}
\end{align}
Here, the derivatives operate on a hard-scattering kernel that has been expanded in transverse momenta, where we define $\bar \omega_1 = n \cdot k_1$ and $\omega_2 = \bar n \cdot k_2$. Using the simplified exponential model for baryon wave functions, we derive the following expressions for the relevant linear combinations appearing in the momentum-space projector: 
\begin{align}
&\chi_3^{(0)}(\bar\omega_1,\omega_2) - \chi_Y(\bar\omega_1,\omega_2) = \frac{\bar\omega_1 \omega_2}{\omega_0^4} \, 
e^{-(\bar \omega_1+\omega_2)/\omega_0} \,, \\
&\chi_3^{(0)}(\bar\omega_1,\omega_2) + \chi_Y(\bar\omega_1,\omega_2) + \chi_{3}^{(i)}(\bar\omega_1,\omega_2)+ \chi_{3}^{(ii)}(\bar\omega_1,\omega_2) = \frac{1}{\omega_0^2} \, 
e^{-(\bar \omega_1+\omega_2)/\omega_0} \,, \\
&\bar\chi_{3}^{(i)}(\bar\omega_1,\omega_2) = \frac{\bar \omega_1 \omega_2}{\omega_0^3} \, 
e^{-(\bar \omega_1+\omega_2)/\omega_0}  \,, \\
&\bar\chi_{3}^{(i)}(\bar\omega_1,\omega_2) + 2 \, \bar\chi_Y(\bar\omega_1,\omega_2) = \frac{\bar \omega_1}{\omega_0^2} \, 
e^{-(\bar \omega_1+\omega_2)/\omega_0}  \,, \\
&\hat\chi_{3}^{(ii)}(\bar\omega_1,\omega_2) = \frac{\bar\omega_1 \omega_2}{\omega_0^3} \, 
e^{-(\bar \omega_1+\omega_2)/\omega_0}   \,, \\
&\hat\chi_{3}^{(ii)}(\bar\omega_1,\omega_2) + 2 \, \hat\chi_Y(\bar\omega_1,\omega_2) = \frac{\omega_2}{\omega_0^2} \, 
e^{-(\bar \omega_1+\omega_2)/\omega_0}  \,. 
\end{align} 

We should note that the arguments $\bar\omega_{1}$ and $\omega_{2}$ in the generalized distribution amplitude are generally not positive definite. This is because the definition of generalized distribution amplitudes requires introducing two Wilson lines with different light-cone directions. The renormalization group evolution kernel of these amplitudes contains a mixing pattern that couples positive and negative supports, thereby extending the support region to the entire real axis\cite{Qin:2022rlk,Huang:2023jdu}.

The matrix $\tilde M^{(2)}$, whose explicit form was obtained in Ref.~\cite{Feldmann:2023plv}, follows the same treatment: 
\begin{align}
    M^{(2)}(\bar{\omega}_1,\omega_2) 
    & =   
    \frac{\slashed n}{2} \, 
    \left(
        \chi_2(\bar{\omega}_1,\omega_2)
        +
        \chi_{42}^{(i)}(\bar{\omega}_1,\omega_2)
    \right)
    \cr
    & +
    \frac{\slashed {\bar n}}{2} 
    \left( 
        \chi_2(\bar{\omega}_1,\omega_2)
        +
        \chi_{42}^{(ii)}(\bar{\omega}_1,\omega_2) 
    \right)
    \cr
    & - 
    \frac{1}{2} \, \gamma_\rho^\perp
    \left( 
        \bar\chi_{42}^{(i)}(\bar{\omega}_1,\omega_2) - \bar\chi_X(\bar{\omega}_1,\omega_2) 
    \right)
    \frac{\slashed n\slashed {\bar n}}{4}
    \, \frac{\partial}{\partial k_{1\rho}^\perp}
    \cr 
    & -
    \frac{1}{2} \, \gamma_\rho^\perp
    \left( 
        \bar\chi_{42}^{(i)}(\bar{\omega}_1,\omega_2) + \bar\chi_X(\bar{\omega}_1,\omega_2) 
    \right)
    \frac{\slashed {\bar n}\slashed n}{4}
    \, \frac{\partial}{\partial k_{1\rho}^\perp}
    \cr 
    & - 
    \frac{1}{2} \, \gamma_\rho^\perp 
    \left( 
        \hat\chi_{42}^{(ii)}(\bar{\omega}_1,\omega_2) + \hat\chi_X(\bar{\omega}_1,\omega_2) 
    \right)
    \frac{\slashed {\bar n}\slashed n}{4}
    \, \frac{\partial}{\partial k_{2\rho}^\perp}
    \cr
    & -
    \frac{1}{2} \, \gamma_\rho^\perp
    \left( 
        \hat\chi_{42}^{(ii)}(\bar{\omega}_1,\omega_2) - \hat\chi_X(\bar{\omega}_1,\omega_2) 
    \right)
    \frac{\slashed n\slashed {\bar n}}{4}
    \, \frac{\partial}{\partial k_{2\rho}^\perp} \,,
    \label{eq:newlcproj2}
\end{align}

and
\begin{align}
\chi_2(\bar\omega_1,\omega_2) + \chi_{42}^{(i)}(\bar\omega_1,\omega_2) &= \frac{\omega_2}{\omega_0^3} \, 
e^{-(\bar \omega_1+\omega_2)/\omega_0} \,, \\
\chi_2(\bar\omega_1,\omega_2) + \chi_{42}^{(ii)}(\bar\omega_1,\omega_2) &= \frac{\bar \omega_1}{\omega_0^3} \, 
e^{-(\bar \omega_1+\omega_2)/\omega_0} \,, \\
\bar\chi_{42}^{(i)}(\bar\omega_1,\omega_2) - \bar\chi_X(\bar\omega_1,\omega_2) &= \frac{\bar \omega_1 \omega_2}{\omega_0^3} \, 
e^{-(\bar \omega_1+\omega_2)/\omega_0}  \,, \\
\bar\chi_{42}^{(i)}(\bar\omega_1,\omega_2) + \bar\chi_X(\bar\omega_1,\omega_2) &= \frac{\bar \omega_1}{\omega_0^2} \, 
e^{-(\bar \omega_1+\omega_2)/\omega_0}  \,, \\
\hat\chi_{42}^{(ii)}(\bar\omega_1,\omega_2) + \hat\chi_X(\bar\omega_1,\omega_2) &= \frac{\omega_2}{\omega_0^2} \, 
e^{-(\bar \omega_1+\omega_2)/\omega_0}   \,, \\
\hat\chi_{42}^{(ii)}(\bar\omega_1,\omega_2) - \hat\chi_X(\bar\omega_1,\omega_2) &= \frac{\bar\omega_1\omega_2}{\omega_0^3} \, 
e^{-(\bar \omega_1+\omega_2)/\omega_0}  \,. 
\end{align} 

\section{Decay distribution in the helicity formalism}
\label{App: helicity amplitude}

In this Appendix, we derive the branching ratios for the decays $\Lambda_b\to\Lambda \, \ell^+\ell^-$ and $\Lambda \, \gamma$ in terms of the effective form factor $\xi_\Lambda(n\cdot p)$.  Following Ref. \cite{Das:2018sms}, we derive the $\Lambda_b\to\Lambda \, \ell^+\ell^-$ decay angular distribution in the helicity formalism. Starting from  the helicity amplitudes defined in \refeq{ SCET-I factorization formulae},   we can further write it as
\begin{align}
\mathcal{M}(s,s',s_1,s_2) &= - \, \xi_\Lambda(n \cdot p) \sum_\lambda \eta_\lambda \bigg[ H_V^\lambda (s, s')L_V^\lambda (s_1,s_2) + H_A^\lambda (s, s')L_A^\lambda (s_1,s_2) \bigg]
 \,.
\label{helicity amplitude}
\end{align}
Here $\lambda=t, \pm1, 0$ are the polarization states of the virtual gauge boson that decays into dilepton pair, $\eta_t = 1$ and $\eta_{\pm1,0} = -1$. $H_{V(A)}^\lambda$ and $L_{V(A)}^\lambda$ are the hadronic and leptonic helicity amplitudes corresponding the vector$(V)$ and axial-vector$(A)$ lepton currents. 

The hadronic helicity amplitudes are the projections of $\Lambda_b\to\Lambda$ matrix elements on the direction of the polarization of virtual gauge boson,
\begin{align}\label{eq:Hl}
H^\lambda_V (s,s') =& \; \bar{\epsilon}^\ast_\mu(\lambda)  \bigg[ C_1^A(E) \, \bar u_\Lambda(p,s') \, (1+\gamma_5) \, \gamma_{\perp}^{\mu} \, u_{\Lambda_b}(P,s) \nonumber \\  &+ C_2^A(E) \, \bar u_\Lambda(p,s') \, (1+\gamma_5) \, u_{\Lambda_b}(P,s) \, \bar n^\mu \bigg] \, ,\\
H^\lambda_A (s,s') =& \; \bar{\epsilon}^\ast_\mu(\lambda)  \bigg[ C_3^A(E) \, \bar u_\Lambda(p,s') \, (1+\gamma_5) \, \gamma_{\perp}^{\mu} \, u_{\Lambda_b}(P,s) \nonumber \\ &+ C_4^A(E) \, \bar u_\Lambda(p,s') \, (1+\gamma_5) \, u_{\Lambda_b}(P,s) \, \bar n^\mu \bigg] \,.
\label{eq:HTll}
\end{align}
Similarly, the leptonic helicity amplitudes are 
\begin{align}\label{eq:Ldef1}
& L^{\lambda}_V = \bar{\epsilon}_\nu(\lambda) \, \bar u(p_1,s_1) \, \gamma^{\nu} \, v(p_2,s_2)\, , \\
& L^{\lambda}_A = \bar{\epsilon}_\nu(\lambda) \, \bar u(p_1,s_1) \, \gamma^{\nu} \, \gamma_5 \, v(p_2,s_2)\,.
\label{eq:Ldef2}
\end{align}
Here $\bar{\epsilon}_\nu(\lambda)$ denotes the polarization vector of the virtual gauge boson. Our choice for the polarization vectors are the same as Ref. \cite{Das:2018sms}. The spinor matrix elements for different combinations of spin orientations are collected in Appendix D of Ref.\cite{Das:2018sms}. Neglecting the mass of the leptons, the leptonic helicity amplitudes can be found in Eq.(B.3)\footnote{It should be notice that the definition of $\bar{\epsilon}^\mu(\pm)$ in Ref.\cite{Das:2018sms} and ${\epsilon}^\mu(\pm)$ in Ref.\cite{Yan:2019tgn} are different with a minus sign, so some results in Eq.(B.3) of Ref.\cite{Yan:2019tgn} need to change with a minus sign.} of Ref.\cite{Yan:2019tgn}. Using these results we can write down the expressions of the helicity amplitudes $\mathcal{M}(s,s',s_1,s_2)$ directly. Finally we get
\begin{align}
\frac{d \Gamma(\Lambda_b \to \Lambda \, \ell^{+} \ell^{-})}{d q^2 \, d \cos \theta} &= \frac{\lambda}{512 \, M_{\Lambda_b}^3 \, \pi^3} \, \times \, \frac{1}{2} \, \sum_s \sum_{s'} \sum_{s_1} \sum_{s_2} \, \left |\mathcal{M}(s,s',s_1,s_2) \right |^2
\nonumber \\
&={\cal N} \left [ F_T(q^2) \left ( 1 + \cos^2 \theta \right )
+ 2 \, F_A(q^2) \, \cos \theta + 2 \, F_L(q^2) \left ( 1 - \cos^2 \theta \right ) \, \right ] ,
\end{align}
where in the factorization limit the helicity amplitudes can be computed as
\begin{eqnarray}
F_T(q^2) &=& \left |\xi_\Lambda(n \cdot p) \right |^2 \, \lambda \, (s_+ + s_-) \, q^2 \, (\left |C_{1}^A\right |^2 + \left |C_{3}^A\right |^2) \,, 
\\
F_A(q^2) &=& 4 \, \left |\xi_\Lambda(n \cdot p) \right |^2 \, \lambda^2 \, q^2 \, {\rm Re} \left({C_1^A}^{\ast} C_3^A \right) \,, 
\\
F_L(q^2) &=& \frac{1}{2} \, \left |\xi_\Lambda(n \cdot p) \right |^2 \, \lambda \, (s_+ + s_-) \, \left |n(q^2) \right |^2 \, (\left |C_{2}^A\right |^2 + \left |C_{4}^A\right |^2) \,, 
\end{eqnarray}
with
\begin{eqnarray}
 {\cal N} &=& \frac{1}{128 \,\, M_{\Lambda_b}^3 \, \pi^3 } \,  \,,   \qquad 
 n(q^2)=\frac{M_{\Lambda_b}^2-M_{\Lambda}^2+q^2+ \lambda}{2 \, M_{\Lambda_b}} \,, 
 \nonumber  \\
\lambda &\equiv& \sqrt{s_+ s_-} = \sqrt{((M_{\Lambda_b}+m_\Lambda)^2-q^2)\,(M_{\Lambda_b}-m_\Lambda)^2-q^2)} \,.  
 \end{eqnarray}
Here, the form factor is $\xi_\Lambda(n \cdot p, \mu)$ and the Wilson coefficients $C_i^A$ have been resummed to LL precison as $C_i^A(n\cdot p, \mu)$. The detailed expressions for  $ C_{i}^{A}(n\cdot p, \mu_h)$ can be found in Ref.\cite{Ali:2006ew}.

For $\Lambda_b\to\Lambda \, \gamma$ decay, we have 
\begin{equation}
\mathcal{M}(s,s',\lambda)=-\, \epsilon_\mu^*(q,\lambda)\, C^A_{\gamma}(n\cdot p)\, \xi_{\Lambda}(q^2=0)\, \bar u_\Lambda(p,s') \,(1+\gamma_5)\, \gamma_{\perp}^{\mu} \, u_{\Lambda_b}(v,s) \,.
\end{equation}
As shown in Ref.\cite{Becher:2005fg}, we have
\begin{equation}
C^A_{\gamma}(\mu)=\frac{G_F}{\sqrt{2}} V_{cs}^* V_{cb} \left[C_7^{\rm eff}+\frac{C_F \alpha_s}{4\pi} \Big( C_8^{\rm eff} G_8+\bar C_2 G_1(x_c) \Big) \right](\mu_{\rm QCD}) \, \Delta_7 C^A (\mu_{\rm QCD}, \mu) +{\cal O}(\alpha_s^2)\,.
\label{eq:CgammaA}
\end{equation}
In practice, we will take $\mu=\mu_{\rm QCD}=\mu_h$. The expressions for $G_1(x), G_8$ and $\Delta_7 C^A$ have been collected in Ref.
\cite{Becher:2005fg}. The RG equation for $C^A_{\gamma}$ is the same as that for the coefficients of the A-type currents in the $\Lambda_b \to \Lambda \, \ell^{+} \ell^{-}$ case, so
\begin{align}
 C_{\gamma}^A(\mu) = \left(\frac{M_{\Lambda_b}}{\mu_h} \right)^{a(\mu_h, \, \mu)} 
{\rm exp} \left [ S(\mu_h, \, \mu) + \frac{5 C_F}{2 \beta_0}\, \ln r_1 \right ] \, C_{\gamma}^A(\mu_h) \,.
\end{align}
Ultimately, we can derive the decay branching ratio as:
\begin{align}
    \mathcal{B}(\Lambda_b \rightarrow \Lambda \gamma)&=\frac{\tau_{\Lambda_b}}{16 \pi M_{\Lambda_b}}\, \left(1-\frac{m^2_{\Lambda}}{M_{\Lambda_b}^2} \right) \times \frac{1}{2}\sum_{s,s'}\sum_{\lambda=\pm}\, \left |\mathcal{M}(s,s',\lambda) \right |^2 \nonumber \\
    &=\frac{\tau_{\Lambda_b}}{4 \pi M_{\Lambda_b}^3} \left(M_{\Lambda_b}^4-m^4_{\Lambda} \right) \left |C_{\gamma}^A(\mu) \right |^2 \left |\xi_{\Lambda}(q^2=0) \right |^2 \,.
    \label{eq:branchingratio}
\end{align}

\newpage

\bibliography{references}
\bibliographystyle{JHEP}

\end{document}

%% file: macros.tex
\newcommand{\refeq}[1]{Eq.~(\ref{eq:#1})}

\renewcommand{\Im}{\text{Im}\,}

\newcounter{TODO}

\makeatletter
    
    \def\tf{\@ifstar\@@tf\@tf}
    \newcommand{\@tf}[1]{{\color{purple}{[\textbf{TF:} #1]}}}
    \newcommand{\@@tf}[1]{{\color{purple}{#1}}}

    \def\ng{\@ifstar\@@ng\@ng}
    \newcommand{\@ng}[1]{{\color{ForestGreen}{[\textbf{NG:} #1]}}}
    \newcommand{\@@ng}[1]{{\color{ForestGreen}{#1}}}
\makeatother